\documentclass[twocolumn]{article}
\usepackage{times}
\usepackage{graphicx}
\usepackage{url}
\usepackage{siunitx}
\usepackage{amsmath}
\usepackage{subcaption}
\usepackage{amsfonts}
\usepackage{gensymb}
\usepackage{braket}
\usepackage{mwe}
\usepackage{xcolor}

\usepackage{float}
\usepackage{algorithm}
\usepackage{algorithmic}
\usepackage{array}

\usepackage{inconsolata} 
\usepackage{booktabs}
\usepackage[numbers,sort&compress,super]{natbib}
\usepackage{textcomp}
\usepackage{tikz}
\usetikzlibrary{shapes.geometric, arrows}

\tikzstyle{startstop} = [rectangle, rounded corners, 
minimum width=3cm, 
minimum height=1cm,
text centered, 
draw=black, 
fill=red!30]

\tikzstyle{io} = [trapezium, 
trapezium stretches=true, % A later addition
trapezium left angle=70, 
trapezium right angle=110, 
minimum width=3cm, 
minimum height=1cm, text centered, 
draw=black, fill=blue!30]

\tikzstyle{process} = [rectangle, 
minimum width=5cm, 
minimum height=1cm, 
text centered, 
text width=5cm, 
draw=black, 
fill=orange!30]

\tikzstyle{decision} = [diamond, 
minimum width=3cm, 
minimum height=1cm, 
text centered, 
draw=black, 
fill=green!30]
\tikzstyle{arrow} = [thick,->,>=stealth]

\usetikzlibrary{positioning}
\newcounter{atomicnumber}
\tikzset{%
  element/.pic={%
    \tikzset{%
      elements/.cd,
      #1,
      /tikz/.cd,
    }%
    \stepcounter{atomicnumber}%
    \node (\elementsymbol) [font=\huge\elementfont, text=\elementtext, inner sep=.5*\elementsep, anchor=mid, fill=\elementfill, rounded corners=2pt, minimum size=\elementsize] {\strut\elementsymbol};
    \node [font=\tiny\elementfont, text=\elementtext, inner sep=2pt, anchor=north west] at (\elementsymbol.north west) {\theatomicnumber};
    \node [font=\tiny\elementfont, text=\elementtext, inner sep=2pt, anchor=south] at (\elementsymbol.south) {\elementname};
  },
  elements/.search also={/tikz},
  elements/.cd,
  name/.store in=\elementname,
  font/.store in=\elementfont,
  text/.store in=\elementtext,
  fill/.store in=\elementfill,
  symbol/.store in=\elementsymbol,
  size/.store in=\elementsize,
  sep/.store in=\elementsep,
  name=Full Name,
  font=\sffamily,
  text=white,
  fill=black,
  symbol=Sy,
  size=35pt,
  sep=2.5pt,
}

\title{Machine Learning to Predict Spectral Anisotropy in Valence-to-Core X-ray Emission Spectroscopy}
\author{Charles A. Cardot, John R. Tichenor, Seth M. Shjandemaar, \\Josh J. Kas,  Gerald T. Seidler, John J. Rehr$^*$\\(*) jjr@uw.edu}

\begin{document}

\twocolumn[
    \maketitle
    \begin{center}
        \begin{minipage}{0.8\textwidth} % Adjust the width of the abstract area
            \raggedright % Left-align the text
            Polarization analysis in x-ray spectroscopy provides an orientation dependent sensitivity to local bonding environments. For a cluster of atoms, polarization sensitivity is most often discussed through the lens of point group symmetries. However, this is a discrete, qualitative method of classifying clusters, and it does little to indicate the degree of spectral anisotropy. Here we adopt a random forest model for quantitative prediction of spectral anisotropy. Its input relies on simplified local geometric and chemical information that can be obtained from any crystal structure file. The model is trained on approximately 11,500 experimentally realized transition metal structures from the Materials Project, with the target being VtC-XES calculated using the real space Green's function code FEFF. We find that the model can strongly predict the degree of spectral anisotropy, with the primary factors being derived from the spatial moments of ligands in a cluster.
        \end{minipage}
        \vspace{1em} % Adds space after the abstract
    \end{center}
]

\section{Introduction}
\hspace{1em}Anisotropic materials have a multitude of applications in a wide variety of fields, including aerospace engineering, high speed electronics, thermo-, photo-, and piezo-electronics, and more \cite{chen_2022, gong_2017,li_2021}.
These materials exhibit directional dependence in mechanical, electronic and magnetic properties, making them highly desirable for use in a wide variety of devices, including optical components in polarization detectors, high-speed transistors, lightweight load-bearing devices, and energy storage solutions. There are many techniques for measuring the anisotropy of material properties \cite{loudon-1964, cahill-2003, ziman-1972}. However, x-ray spectroscopy and related techniques are unique in their ability to probe the anisotropy of the local atomic, magnetic, and electronic structure of the material, allowing analysis of the fundamental structure-function relationships \cite{tortora-2024}.
Polarization is prevalent in x-ray spectroscopy as a method of analyzing spin and orbital magnetic moments \cite{van-der-laan-2014}, magnetic axes or interfaces \cite{jo-2004, van-der-laan-2011}, crystal field effects \cite{bergmann-2002}, and coordination environments \cite{jansing-2016}. With the advent and improvement of high brilliance synchrotron radiation over the last 40 years, a diverse suite of experimental capabilities have emerged that can use polarization as an additional probe within x-ray spectroscopy to extract electronic and chemical information \cite{glatzel-2004, pollock-2015, ketenoglu-2022, trippe-2014, stohr-1992}. In x-ray absorption spectroscopy (XAS) it can be used as a spin-dependent probe of the unoccupied density of states via x-ray magnetic circular dichroism (XMCD) \cite{stohr-1999}, or as a probe of bond directions and magnetic axes in transition metal oxides through x-ray linear dichroism (XLD) \cite{bianconi-1988}. 

While polarization-dependent effects have been extensively studied in the absorption regime, comparatively fewer studies have investigated polarization in x-ray emission spectroscopy (XES). This is in part due to the relative difficulty of measuring the polarization of emitted light \cite{fumagalli-2019} compared to creating a polarized x-ray source at a synchrotron \cite{kim-2017}. Specifically, for the study of 3$d$ transition metals, valence-to-core XES (VtC-XES) is a powerful element specific probe of the occupied local density of states with sensitivity to ligand identity, oxidation state, and coordination geometry \cite{pollock-2015, nascimento-2022, roemelt-2024}. Unlike core-to-core transitions (ex: K$\alpha$ XES), VtC-XES directly probes the local valence electronic structure, and thus inherently encodes local anisotropy. Another major advantage is that, compared to absorption techniques, VtC-XES avoids the presence of a deep core hole in the final state, and therefore provides a more representative picture of the ground state electronic structure of the system \cite{geoghegan-2022}. It has also been shown that \textit{ab initio} theoretical methods are able to accurately reproduce polarized x-ray emission spectra \cite{bergmann-2002, abramson_x-ray_2025}. For these reasons, we focus on VtC-XES in this work, aiming to quantify and model the anisotropy in emission spectra as a function of local chemical bonding environment.

\begin{figure}
    \centering
    \includegraphics[width=1\linewidth]{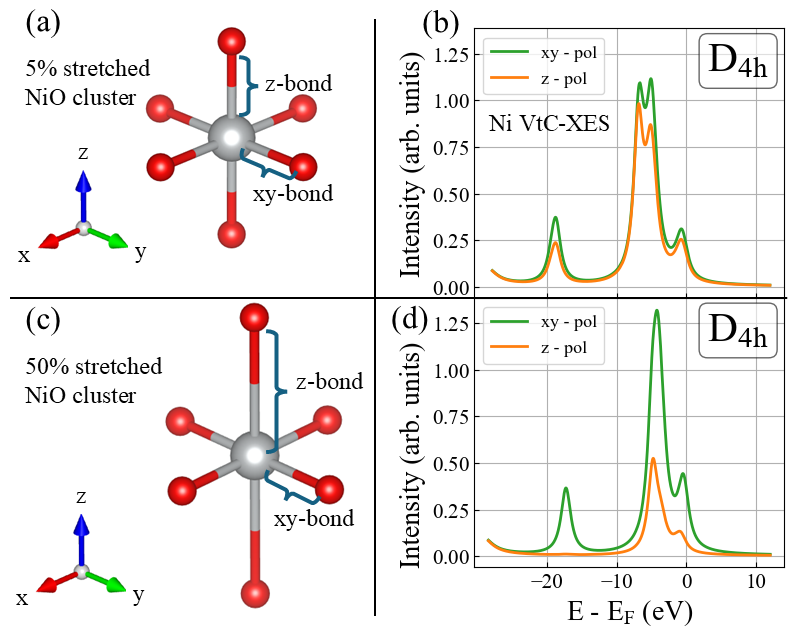}
    \caption{(a, c) Stretched NiO cluster, with the Ni (silver) to O (red) bond along the z-axis stretched 5\% and 50\% respectively. (b, d) The Ni VtC-XES polarized along the xy bonds in green and z bond in orange for the clusters in (a) and (c) respectively.}
    \label{fig:NiO_D4h_stretch.png}
\end{figure}

Local anisotropy is regularly treated in terms of discrete classifications, namely point group symmetries \cite{watanabe-1966, brener-2024, nielsen-2024}. These are used in conjunction with oxidation state to interpret the ground state electron configuration for transition metal oxides in a given crystal field (ex: high-spin and low-spin) \cite{lafuerza-2020}. However, it is difficult to predict \textit{a priori} how large of an effect the point group derived crystal field will have on a general x-ray spectrum. 
This is illustrated in Figure \ref{fig:NiO_D4h_stretch.png}, which shows how the polarization dependent VtC-XES changes when going from a 5\% stretch of the apical O bond-length Figure \ref{fig:NiO_D4h_stretch.png}a,b to a 50\% stretch Figure \ref{fig:NiO_D4h_stretch.png}c,d. While the extent of broken octahedral (Oh) symmetry is vastly different, these clusters have the \textit{same} point group (D$_{4h}$).

Thus, there is a need for a continuous, quantitative metric that encodes the local anisotropic environment in a way that can be used for predicting polarization dependence. Such a metric would enable rapid interpretation of experimental data and support the development of artificial intelligence (AI) driven workflows at synchrotron facilities. AI-based tools have already been successfully applied to phase identification, as well as parameter extraction in x-ray diffraction (XRD) and x-ray reflectometry (XRR), enabling continuous and rapid data-to-decision pipelines for beamline science \cite{pithan-2023}. This metric can also be used for bulk, quantitative screening of large materials databases to identify candidates with strong local anisotropy effects. 
% In particular, it enables efficient discovery of high-spin and low-spin systems by sorting materials based on their spectral anisotropy signatures, which can reflect underlying spin-state configurations. Such anisotropy often correlates with key electronic phenomena such as orbital ordering and spin-state transitions \cite{pollock-2015}. 
This kind of targeted exploration speeds up the search for functional materials in areas like spintronics \cite{zutic-2004, wang-2023, tombros-2008}, quantum information \cite{shen-2022, hirohata-2025}, and energy conversion and storage \cite{sendek-2018}.

% high-throughput materials screening, inverse design of functional optical materials, and interpretation of experimental x-ray spectroscopies \textbf{need more citations to examples}. 

% It would also provide a route to disentangle local non-magnetic contributions to anisotropy from magnetic ones \textbf{need citations}. For example, in Haverkort \textit{et al.} the L$_{2,3}$ dichroism in the Ni L$_{2,3}$ absorption of monolayer NiO films is analyzed using a combination of multiplet ligand field theory and temperature varying magnetic linear dichroism measurements. They prove that, despite the similarity to dichroic signals that arrive from the antiferromagnetic behavior of thicker NiO films, the dichroism from the monolayer is purely due to crystal field effects at low symmetry. Dichroism is a result of a combination of crystal field anisotropy and local spin polarization. A geometric–chemical anisotropy metric like the one we propose here could help isolate and quantify each contribution.

% Recent advances in \textit{ab initio} theoretical methods have provided a reliable way to routinely calculate various x-ray spectroscopies \textbf{cite many}. However, despite their successes, these methods remain computationally intensive and are often limited by system size, core-hole treatment, and the need for fine spectral resolution. 

%Moreover, ML models can learn to interpolate smoothly between different chemistries and structural motifs, potentially capturing subtle trends in the data that might be missed by more rigid physical approximations.

Recent advances in \textit{ab initio} theoretical methods have made it possible to reliably calculate a wide range of x-ray spectroscopies \cite{vinson-2011, apra-2020}. Despite these successes, such calculations remain computationally demanding and are often limited by system size and the need to treat electron correlation \cite{pople-1999}. By contrast, a trained machine learning model can make predictions orders of magnitude faster, enabling rapid exploration of material spaces.

The relationship between a material’s local chemical and geometric environment, and its polarization-dependent spectral response, is inherently nonlinear and highly complex. Spectral anisotropy arises from directional variations in the electronic states, which are influenced by factors such as local bonding asymmetry, ligand field distortions, coordination number, and electronic hybridization \cite{Drager.Brummer1984}. Traditional linear or descriptor-based approaches struggle to capture these coupled, high-dimensional effects. Machine learning models, by contrast, are well-suited to learn nonlinear mappings from structural and chemical features to spectral response.

Machine learning has successfully been applied to several modalities of spectroscopy including x-ray absorption near-edge structure (XANESNet) \cite{rankine-2020}, extended x-ray absorption fine structure \cite{timoshenko-2018}, and vibrational spectroscopy \cite{gastegger-2017}. The broader field of materials sciences has also demonstrated that machine learning models can successfully predict bulk material properties, such as utilizing a message-passing graph neural network (MPGNN) to capture anisotropic properties such as elastic \cite{wen-2024} and dielectric tensors \cite{lou-2024}. However, these GNN approaches can be limited by the scarcity of high-quality training data and, like most neural network-based methods, suffer from a black-box nature that limits interpretability.

In contrast, random forest models have demonstrated strong performance in materials spectroscopy; for example, Zheng et al. successfully applied them to identify coordination environments in XANES spectra \cite{Zheng.etal2020}. Random forests can capture nonlinear and interacting effects among structural and chemical features, making them well-suited for predicting complex spectral responses while maintaining interpretability. To the authors’ knowledge, anisotropy in the VtC-XES regime remains largely unexplored and has not yet been the subject of any machine learning applications. In this paper, we develop a random forest regression (RFR) model trained on structural and chemical descriptors of 3$d$ transition metal complexes to predict the spectral anisotropy in VtC-XES. To quantify the degree of anisotropy, we introduce a new metric termed the spectral anisotropy matrix sum (SAMS).

The rest of this article is organized as follows. In Section \ref{sec:methods} we described the procedure for quantifying the spectral anisotropy and how we extract input features as our predictors in the model. We also address the calculation pipeline used for developing the training data set, and the procedure used to develop the random forest model. In Section \ref{sec:RandD} we present the core results of the model, namely the $R^2$ and mean absolute error (MAE). We also assess the relative importance of the features and compare predictive capabilities across the 10 3$d$ transition metals. Finally in Section \ref{sec:conclusion} we discuss constraints of the model and propose future directions for how it could be improved and extended to new applications.

\section{Methods}
\label{sec:methods}

\subsection{Spectral Anisotropy Matrix}

\hspace{1em}To quantify the degree of anisotropy in a spectrum we introduce a metric based on the $L^2$-norm or Euclidean norm of the difference between individual polarizations \cite{bishop-2007}. We start by decomposing a spectrum into the emission for three orthogonal polarization axes ($\sigma_i, \sigma_j, \sigma_k$). Calculating the square root of the integral over the difference squared, we arrive at a single number to characterize the "difference" between two spectra. This value is normalized by the integral of the average (isotropic) spectra, $\bar{\sigma}$. We calculate this parameter for each possible pair of axes to  produce the spectral anisotropy matrix (SAM) as defined in eq \ref{eq:SAM}.

\begin{align}
    SAM_{ij} &= \frac{\left[ \int _{a} ^b |\sigma_{i}(\epsilon) - \sigma_{j}(\epsilon)|^2 \, d\epsilon \right]^{1/2}}  {\int_a^b \bar{\sigma}(\epsilon) d\epsilon} \label{eq:SAM} \\
    SAM_{sum} &= SAM_{ij} + SAM_{jk} + SAM_{ik} \label{eq:SAMS}
\end{align}

\begin{figure}
    \centering
    \includegraphics[width=0.9\linewidth]{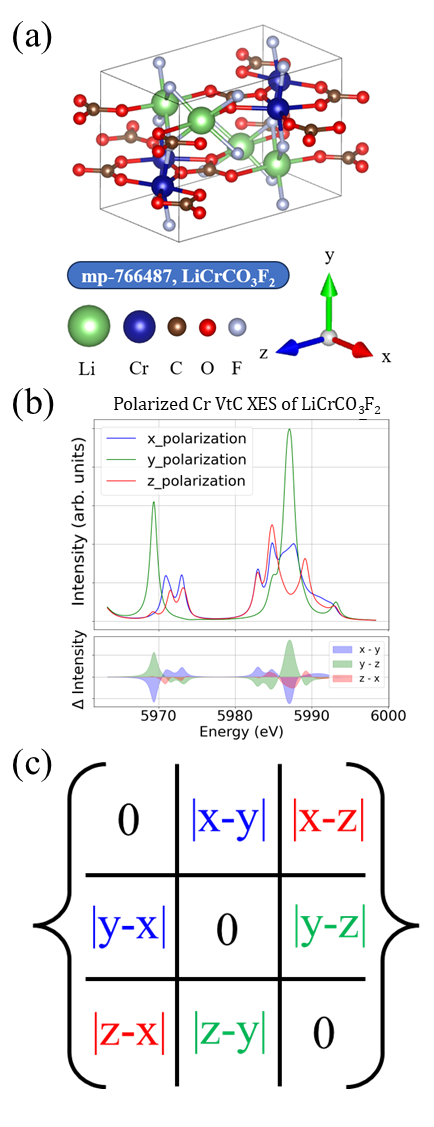}
    \caption{(a) Crystal structure of LiCrCO$_3$F$_2$, where Cr is bonded to five O atoms and one F atom in a distorted O$_h$ cluster. (b) Polarized VtC Cr XES along the $x$ (blue), $y$ (green), and $z$ (red) axes, with differences between each pair shown below. (c) Spectral Anisotropy Matrix (SAM), where diagonal values are zero and off-diagonal values are the Euclidean norm of the differences ($|x-y|$, $|x-z|$, $|y-z|$).}
    \label{fig:SAM_example}
\end{figure}

The SAM is a zero-diagonal symmetric matrix. A visualization of the calculation process is shown in Figure \ref{fig:SAM_example}, where the Cr VtC-XES is calculated along the $x$, $y$, and $z$ crystal axes of an example transition metal crystal system, LiCrCO$_3$F$_2$. The crystal structure is shown in Figure \ref{fig:SAM_example}a, where the local structure around the Cr atoms is distorted such that there are no degenerate axes. The polarized spectra along the three crystal axes is shown in Figure \ref{fig:SAM_example}b along with the difference curves between each pair of spectra. These are then used to calculate the Euclidean norm and build the SAM shown in Figure \ref{fig:SAM_example}c. Taking the sum of the upper off diagonal components as shown in eq \ref{eq:SAMS}, we arrive at a quantitative measure, the spectral anisotropy matrix sum (SAMS). This metric captures the total amount of anisotropy in the spectrum, and is the numerical value we aim to predict with our model in this work. A SAM is highly dependent on the basis used to define the cluster around the transition metal ion. Our goal is to create a dataset of the maximum possible SAMS signal that can be achieved for a given structure, which often requires rotating to the principal axes. 

\subsection{Feature Anisotropy Matrices}

\hspace{1em}To parameterize the anisotropy of the local atomic environment around an absorber atom, we construct 3 related but distinct metrics derived from the quadrupole moment eq \ref{eq:weighted_quad}, dipole moment eq \ref{eq:weighted_dipole}, and inertia tensor eq \ref{eq:weighted_inertia} respectively.

\begin{subequations}\label{eq:weighted_all}
\begin{align}
    Q^{\mathrm{weighted}}_{ij} 
        &= \sum_{n} q_n \, \frac{r_{n,i}\, r_{n,j}}{\lVert \vec{r}_n \rVert^7}  \label{eq:weighted_quad}\\
    D^{\mathrm{weighted}}_{i} 
        &= \sum_{n} \frac{q_n\, r_{n,i}}{\lVert \vec{r}_n\rVert^7} \label{eq:weighted_dipole}\\
    I^{\mathrm{weighted}}_{ij} 
        &= \sum_{n} \frac{r_{n,i}\, r_{n,j}}{\lVert \vec{r}_n \rVert^7} \label{eq:weighted_inertia}
\end{align}
\end{subequations}

Here, \( q_n \) is the effective charge assigned to the \( n \)-th neighboring atom (estimated via oxidation state), \( r_{n,i} \) and \( r_{n,j} \) are the \( i \)-th and \( j \)-th Cartesian components of the position vector \( \vec{r}_n \) relative to the absorber atom, and \( \lVert \vec{r}_n \rVert \) is the Euclidean distance from the absorber to the \( n \)-th atom. To emphasize the influence of nearby atoms, we introduce a distance-dependent weighting factor of $1/\|\vec{r}_n\|^7$, ensuring that contributions from closer atoms dominate while those from more distant atoms decay rapidly. This steep falloff reflects the localized nature of anisotropic interactions relevant to x-ray emission and allows the resulting descriptor to better capture short-range geometric features. The choice of exponent was determined through systematic optimization over integer values ranging from 1 to 14. 

The quadrupole and inertia matrices are identical with the exception of the effective charge $q_n$, which serves to encode chemical information. We will see that the anisotropy metrics of both are strongly correlated with the spectral anisotropy, but we choose to use the eigenvectors of the weighted quadrupole moment to define a rotation-invariant basis. This is done both to maximize the observed spectral anisotropy in eq \ref{eq:SAMS} and provide a set coordinate system to work in. The main reason for including both the quadrupole eq \ref{eq:weighted_quad} and the inertia eq \ref{eq:weighted_inertia} comes from the reliance on estimating the ligand charges $q_{n}$. We rely on simple oxidation state estimations and avoid doing any direct electronic structure calculations for generating input features, and uncertainties in the charge estimates can degrade the reliability of the quadrupole metric. Incorporating a purely geometric metric thus provides a complementary, charge-independent measure of local anisotropy that can help disentangle geometric effects from uncertainties in ligand charge assignment. 

The quadrupole, dipole, and inertia anisotropy matrices (denotes QAM, DAM, and IAM respectively) are shown in eq \ref{eq:AM_all}. They are all expressed in the eigen-basis of the quadrupole matrix (eigenvalues $\lambda$), where $\tilde{D} = V^{-1} D V$ and $\tilde{I} = V^{-1} I V$ are the dipole and inertia matrices transformed into this basis and $V$ is the matrix whose columns are the orthonormal eigenvectors of the weighted quadrupole moment.

\begin{subequations}\label{eq:AM_all}
\begin{align}
    QAM_{ij} &= |\lambda_i - \lambda_j| \label{eq:QAM}\\
    DAM_{ij} &= \lvert \tilde{D}_i - \tilde{D}_j \rvert \label{{eq:DAM}}\\
    IAM_{ij} &= \lvert \tilde{I}_i - \tilde{I}_j \rvert \label{eq:IAM}
\end{align}
\end{subequations}

Consistent with the single anisotropy measure used for summarizing the spectral anisotropy metric into the SAMS in eq \ref{eq:SAMS}, a similar sum is applied to the off diagonal elements of eq \ref{eq:AM_all} to arrive at the quadrupole, dipole, and inertia anisotropy matrix sums, denoted QAMS, DAMS, and IAMS respectively (eq \ref{eq:AMS_all}). If a unit cell contains multiple unique sites for a 3$d$ transition metal, we average the anisotropy matrix sum over all sites to obtain a single representative number for the entire crystal.

\begin{subequations}
\label{eq:AMS_all}
\begin{align}
    QAMS &= QAM_{ij} + QAM_{jk} + QAM_{ik} \label{eq:QAMS} \\
    DAMS &= DAM_{ij} + DAM_{jk} + DAM_{ik} \label{eq:DAMS} \\
    IAMS &= IAM_{ij} + IAM_{jk} + IAM_{ik} \label{eq:IAMS}
\end{align}
\end{subequations}

\subsection{Connection between Quadrupole and Spectra}

\label{sec:Quad_vs_Spec}

% The perturbing Hamiltonian $H$ can be written as $H \approx \hat{\epsilon} \cdot \vec{r}$, where $\hat{\epsilon}$ is the polarization vector of the photon and $\vec{r}$ is a spatial coordinate vector. 

\hspace{1em}Although the IAMS and QAMS are closely related and provide comparable predictive power for the spectral anisotropy, we restrict our analysis here to the QAMS to avoid redundancy. To better understand the effectiveness of this metric it is useful to examine how both the QAMS and SAMS features reflect the underlying physical mechanisms governing anisotropy. The transition rate between an initial state $\ket{i}$ and final state $\bra{f}$ is given by Fermi's Golden rule in eq \ref{eq:FermisGoldenRule}. $H$ is the perturbing Hamiltonian, $\rho_f$ is the projected density of states, $E_f$ and $E_i$ are the energies of the final and initial states respectively, and $\hbar \omega$ is the energy of the emitted x-ray photon \cite{sakurai1967advanced}.

\begin{equation}
    \Gamma_{i \rightarrow f} \propto \frac{2 \pi}{\hbar} |\bra{f}H\ket{i}|^2 \rho_f(E_f - E_i -\hbar \omega)
    \label{eq:FermisGoldenRule}
\end{equation}

The continuous form of the quadrupole moment tensor is defined in eq \ref{eq:continuous_quad} (without additional distance weighting). This gives the quadrupole moment of a charge distribution with electron density \( \rho_e(\vec{r}) \) (eq \ref{eq:electron_density}), where \( e \) is the electron charge, $N$ is the number of electrons, and $\Psi$ is the many-body wavefunction.

\begin{equation}
    Q^{continuous}_{i,j} = -\int e \rho_e(\vec{r}) r_i r_j d^3\vec{r}
    \label{eq:continuous_quad}
\end{equation}

\begin{equation}
    \rho_e(\vec{r}) = N \int |\Psi(\vec{r},\vec{r}_2,...,\vec{r}_N)|^2 d^3\vec{r}_2...d^3\vec{r}_N
    \label{eq:electron_density}
\end{equation}

In eq \ref{eq:FermisGoldenRule}, the initial and final state wavefunctions are shaped by the local chemical environment. This manifests as polarization-dependent effects in the spectrum. Since the electron density in eq \ref{eq:electron_density} is derived from the modulus squared of the many-body wavefunction, it retains similar spatial anisotropy, albeit without phase information. The quadrupole moment is therefore a natural option for encoding this anisotropy, capturing both geometric and electronic symmetry-breaking features of the local environment.

% \begin{align}
%     |\bra{f}H\ket{i}|^2 &= | \braket{f|{\vec{x}_1,\vec{x}_2,...,\vec{x}_N}} \\ &\times \bra{{\vec{x}_1,\vec{x}_2,...,\vec{x}_N}}H\ket{{\vec{y}_1,\vec{y}_2,...,\vec{y}_N}} \notag \\ &\times \braket{{\vec{x}_1,\vec{x}_2,...,\vec{x}_N}|i}|^2 \notag \\
%     &= |\Psi_f|^2 \braket{H} |\Psi_i|^2 \notag \\
%     &= \rho_{e,f}(\vec{r}) \braket{H} \rho_{e,i}(\vec{r}) \notag
% \end{align}

\begin{figure}
    \centering
    \includegraphics[width=\linewidth]{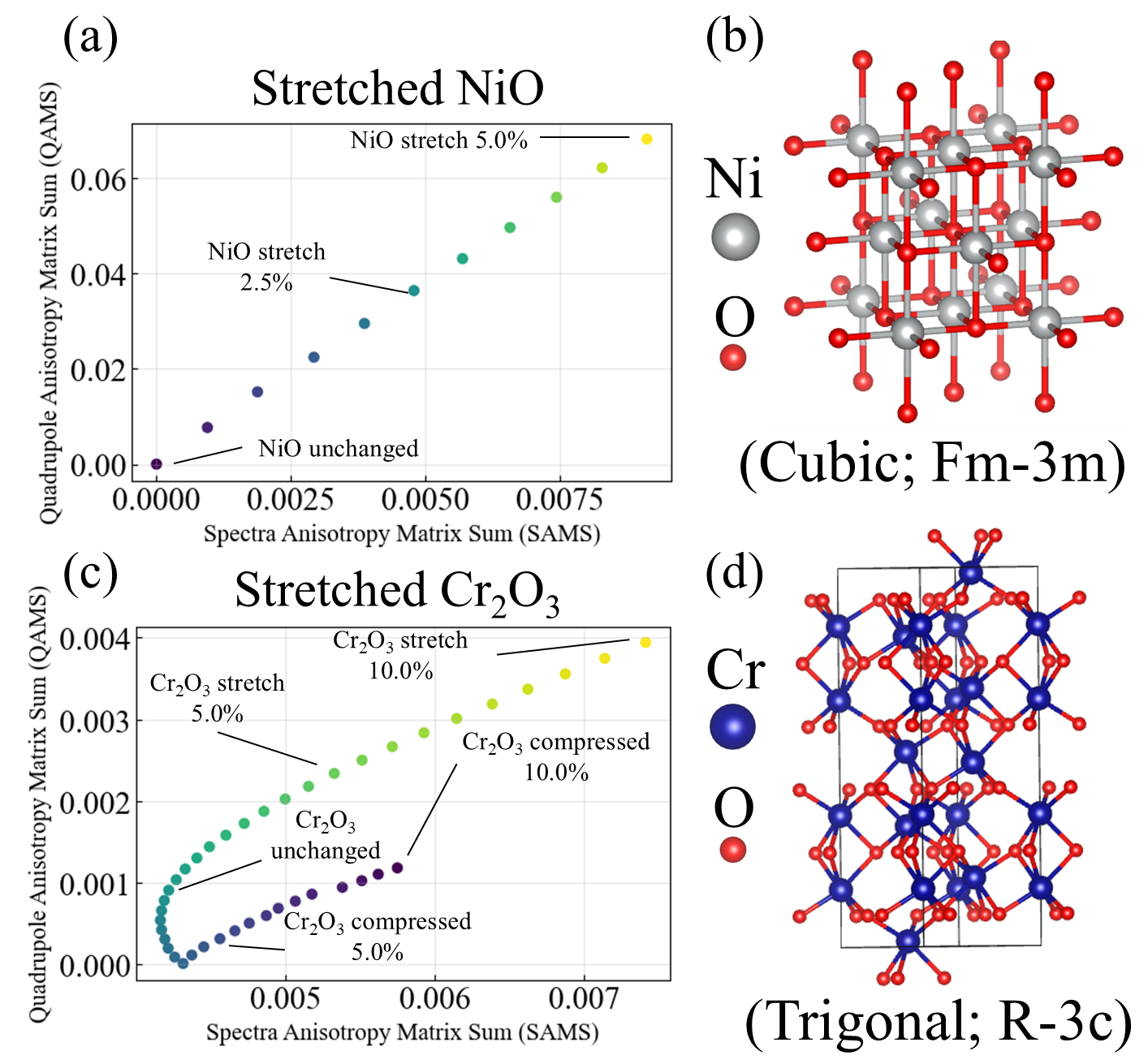}
    \caption{QAMS versus SAMS parameters for distorted crystal structures, NiO (a, b) and Cr$_2$O$_3$ (c, d). NiO exhibits a linear relationship between the two parameters in response to stretching, but Cr$_2$O$_3$ has a nonlinear response the in the region of small distortions around the base experimental structure.}
    \label{fig:QAMS_vs_SAMS_motivation}
\end{figure}

To further assess the relationship between the QAMS and the SAMS, we now examine how they respond to systematic geometric distortions in representative materials, specifically NiO and Cr$_2$O$_3$, as shown in Figure \ref{fig:QAMS_vs_SAMS_motivation}. Figure \ref{fig:QAMS_vs_SAMS_motivation}b,d show the crystal structure for NiO and Cr$_2$O$_3$ respectively, while Figure \ref{fig:QAMS_vs_SAMS_motivation}a,c show the QAMS vs. SAMS for different degrees of stretch or compression from the experimental structure. In Figure \ref{fig:QAMS_vs_SAMS_motivation}a the NiO structure is stretched along the $c$-axis in increments of 0.5\%. Starting from the lower left-hand corner of the plot—where all bond lengths are equal—the structure is isotropic, and both QAMS and SAMS are exactly zero, which is in agreement with the perfect cubic NiO structure. As the structure becomes increasingly stretched, both QAMS and SAMS vary linearly. This suggests that for simple distortions such as bond elongation without significant angular rearrangement, QAMS tracks spectral anisotropy with high fidelity.

% \begin{figure}[h]
%     \centering
%     \includegraphics[width=1\linewidth]{figures/QAMSvSAMSNiO.jpeg} 
%     \caption{QAMS vs. SAMS for NiO under uniaxial stretching along the $c$-axis in 0.5 \AA\ increments. A linear relationship is observed.}
%     \label{fig:niO_qams_sams}
% \end{figure}

However, in the case of a similar incremental distortion of Cr$_2$O$_3$ in Figure \ref{fig:QAMS_vs_SAMS_motivation}c we observe a distinctly nonlinear relationship between QAMS and SAMS. Here the local environment around the Cr ion is a distorted O$_h$ structure, and has non-zero QAMS and SAMS for even the starting crystal structure. In the extremes of both compression and stretching, the relationship between the two parameters remains linear. In the region around the unchanged structure, we note that the relationship deviates from linearity, and that the minimum of the SAMS does not line up with the minimum of the QAMS. 

This behavior leads to two key insights. First, the nonlinear behavior arises from changes not only in bond lengths but also in bond angles, which substantially affect the spatial charge distribution and, consequently, the spectral anisotropy. For NiO, the stretching occurs uniformly without altering angular geometry, but Cr$_2$O$_3$ has a more complex structural response to distortion. While the two parameters are clearly correlated, their relationship is not strictly proportional. Second, it is possible to have a local cluster that produces a zero QAMS and a non-zero SAMS (for example, a perfectly linear molecule where the ligands have the same charge $q$). The DAMS partially addresses this limitation by capturing dipolar contributions to the anisotropy, as it is derived from a metric that is odd under spatial inversion. However, this alone is insufficient, and additional features are required to more fully capture and discriminate between the underlying sources of spectral anisotropy.

\subsection{Additional Input Features}

\begin{figure}[h]
    \centering
    \includegraphics[width=1\linewidth]{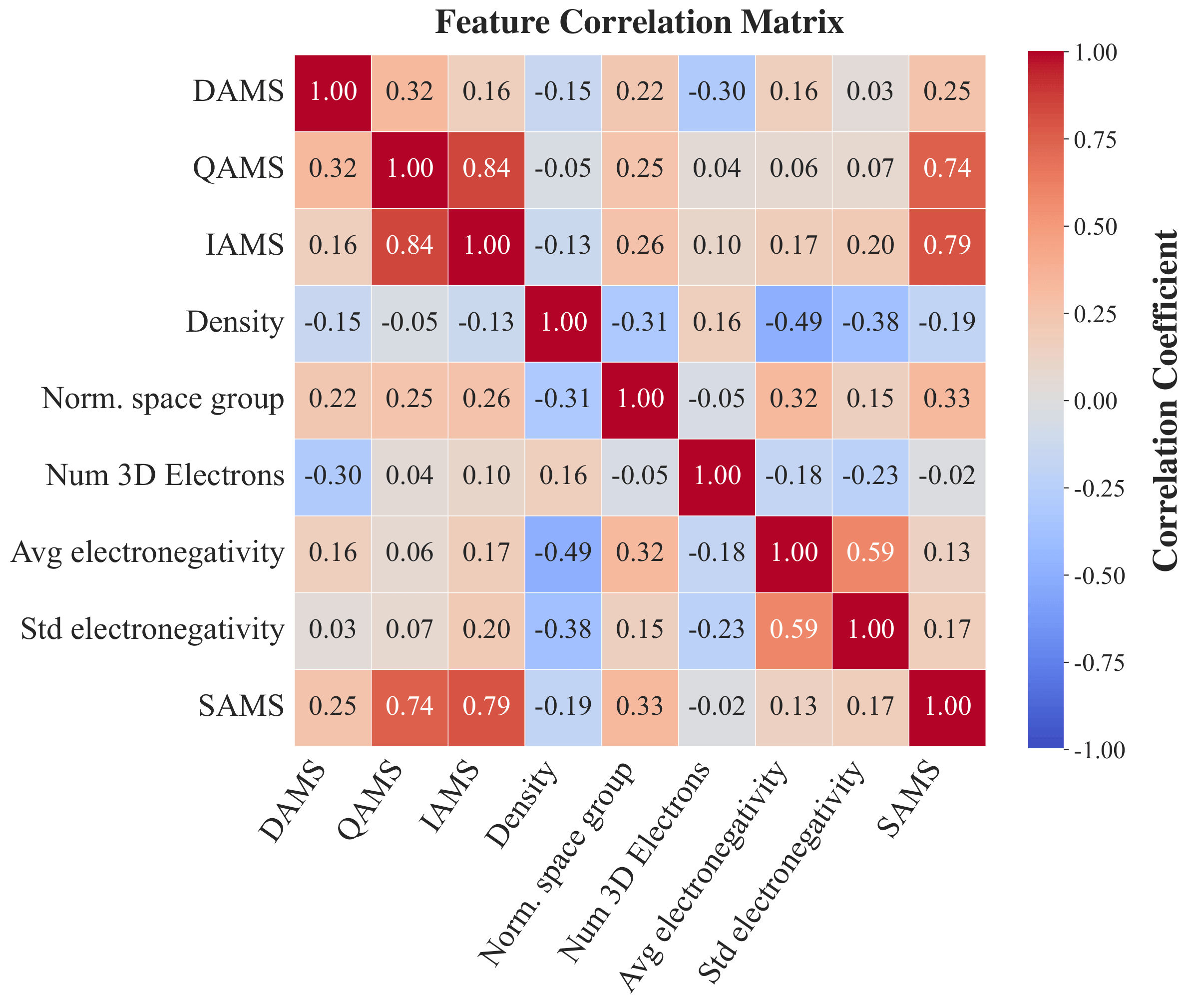}
    \caption{Correlation matrix of the features. Each cell shows the Pearson correlation coefficient between pairs of features, with values ranging from $-1$ (perfect anticorrelation, blue) to $+1$ (perfect correlation, red).}
    \label{fig:factor_correlation_matrix}
\end{figure}

\hspace{1em}This section focuses on defining each additional descriptor beyond the anisotropy-matrix–derived quantities and motivating its inclusion in the model. As we have shown, the QAMS alone is not a strong predictor of the SAMS, which motivates incorporating additional local geometric and chemical factors. Along this line, we include the mass density, the normalized space group number, the number of 3$d$-electrons, average and standard deviation of the electronegativity of the cluster as additional factors. The correlation between all factors, along with the SAMS, is displayed in Figure \ref{fig:factor_correlation_matrix}. 

The normalized space group number ($N_{SG}$), as shown in eq \ref{eq:Norm_space_group},
\begin{equation}
    N_{\mathrm{SG}} = 1 - \frac{N_s}{N_t}
    \label{eq:Norm_space_group}
\end{equation}
where $N_s$ is the space group number of the structure, and $N_t = 230$ is the total number of crystallographic space groups. Equation \ref{eq:Norm_space_group} provides a rough, simple analogue for the symmetry of the crystal system and, to a lesser extent, the symmetry of the local environment around the metal.

The number of $3d$ electrons on the central atom is given in eq \ref{eq:num_3d},

\begin{equation}
    N_{3d} = Z_c - 18 - q_c,
    \label{eq:num_3d}
\end{equation}

where $Z_c$ is the atomic number of the central atom and $q_c$ is its charge, as determined by the Pymatgen python library oxidation state guesser. The constant $18$ accounts for the electrons in the filled inner shells (\(1s^2\,2s^2\,2p^6\,3s^2\,3p^6\)), and this parameter is intended to add additional chemical information in a way that is analogous to oxidation state.

The average and standard deviation of electronegativity were added as features to capture the distribution of atomic bonding tendencies within each cluster. These quantities were computed using the Pauling electronegativity values provided by the Pymatgen periodic table \cite{Ong.etal2013}. For each cluster, we evaluated the electronegativity of all constituent atoms and then calculated the mean and standard deviation. The mass density of the structure is obtained from the Materials Project \cite{Jain2013}.

Overall, Figure \ref{fig:factor_correlation_matrix} indicates that while none of the additional descriptors individually shows a dominant linear correlation with the SAMS, several exhibit weak-to-moderate correlations that are distinct (uncorrelated) from those of the matrix-based anisotropy quantities.

% Although features such as number of 3d electrons, standard and average electronegativity show little direct linear correlation with the SAMS in figure 4, it has been shown that including factors correlated with other descriptors \cite{janet-2017}, rather than only with the target value, can improve machine learning model performance. There may also be complex non linear relationships between these features and the SAMS that the Pearson's correlation matrix cannot find.

% \subsubsection{Feature Calculation Pipeline}

% The features are computed as follows. From a crystallographic information file (CIF), we construct the local environment including up to 12 \AA around around each unique $3d$ transition metal site. For each unique site, we compute $Q$, $I$, and $D$. These quantities are then averaged over all unique sites, and the averaged $Q$, $I$, and $D$ are used to compute the $QAMS$, $IAMS$, and $DAMS$. The $N_\text{SG}$ and $N_{3d}$ are computed only once, as these quantities do not vary between sites. The density of the structure is obtained from the Materials Project \cite{Jain2013}.

\subsection{Dataset curation}

\begin{figure}
    \centering
    
    \includegraphics[width=1\linewidth]{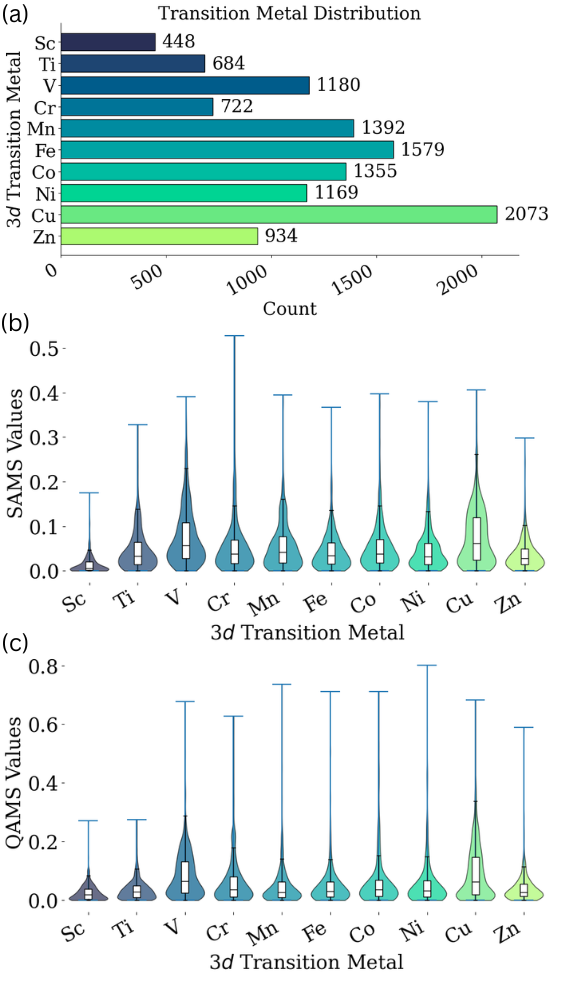} 
    \caption{(a) Distribution of the 3$d$ transition metals in the dataset, showing the total number of structures containing each metal. (b, c) Distribution of (SAMS, QAMS) grouped by transition metal.}
    \label{fig:TM_dataset_distribution}
\end{figure}

Crystallographic Information Files (CIFs) were queried from the Materials Project in their primitive cell format \cite{Jain2013}. We selected structures that are experimentally observed and that contain at least one 3$d$ transition metal atom, resulting in a dataset of approximately 11,500 compounds. Figure \ref{fig:TM_dataset_distribution}a shows the distribution of transition metals within the dataset, illustrating that certain elements such as Cu and Fe are overrepresented relative to others like Sc or Ti, reflecting trends in the Materials Project database. Figures \ref{fig:TM_dataset_distribution}b,c further show the distributions of the QAMS and SAMS metrics respectively. While not identical for every element, the QAMS and SAMS distributions for the same transition metal are qualitatively similar. The most spectrally isotropic element is Sc, with very few compounds exhibiting any significant anisotropy. As we will see in Section \ref{sec:RandD}, the limitations of the Sc subset (both the limited number of compounds and the narrow distribution) make it difficult for the model to sufficiently generalize to all 3$d$ transition metals. When splitting the dataset into training and testing subsets, this elemental imbalance is mitigated by selecting a proportionally representative number of structures for each transition metal.

\subsection{Corvus Workflow and FEFF}

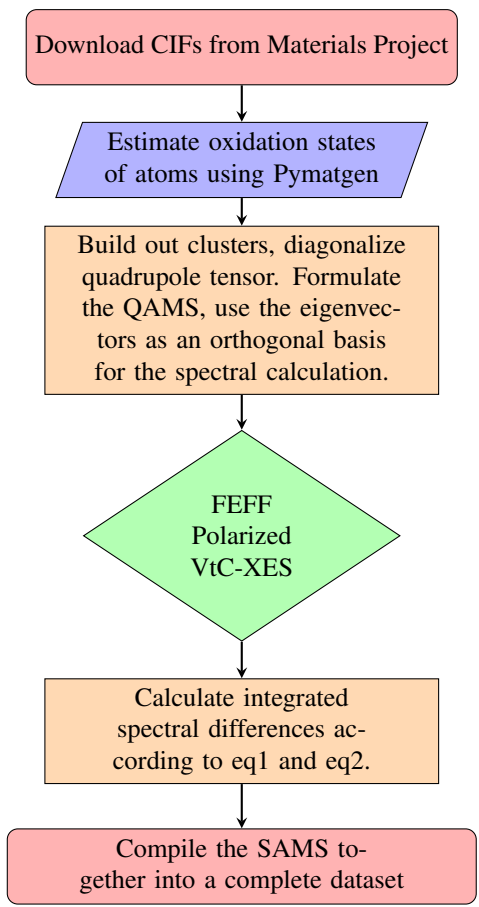
\begin{figure}[h]
    \centering

    \caption{Overview diagram of the Corvus workflow used to construct the dataset. Materials were obtained from the Materials Project and restricted to compounds containing at least one 3$d$ transition-metal ion that are flagged as experimentally observed.}
    \label{fig:Corvus-workflow}

\begin{tikzpicture}[node distance=2cm]

\node (start) [startstop] {Download CIFs from Materials Project};
\node (in1) [io, below of=start, text width=4cm, yshift=0.5cm] {Estimate oxidation states of atoms using Pymatgen};
\node (pro1) [process, below of=in1] {Build out clusters, diagonalize quadrupole tensor. Formulate the QAMS, use the eigenvectors as an orthogonal basis for the spectral calculation.};
\node (dec1) [decision, below of=pro1, yshift=-1.0cm, text width=2cm, aspect=1.5] {FEFF Polarized VtC-XES};
\node (pro2a) [process, below of=dec1, yshift=-0.6cm] {Calculate integrated spectral differences according to eq\ref{eq:SAM} and eq\ref{eq:SAMS}.};
%\node (pro2b) [process, right of=dec1, xshift=2cm] {Adjust SCF and FMS radius};
\node (stop) [startstop, below of=pro2a, text width=6cm, yshift=0.2cm] {Compile the SAMS together into a complete dataset};

\draw [arrow] (start) -- (in1);
\draw [arrow] (in1) -- (pro1);
\draw [arrow] (pro1) -- (dec1);
\draw [arrow] (dec1) -- (pro2a);
%\draw [arrow] (dec1) -- node[anchor=east] {convergence} (pro2a);
%\draw [arrow] (dec1) -- node[above=5mm] {non-convergence} (pro2b);
%\draw [arrow] (pro2b) |- (pro1);
\draw [arrow] (pro2a) -- (stop);

\end{tikzpicture}
\end{figure}

Corvus is an automation workflow handler \cite{Kas.etal2021a} which uses a CIF and user defined information to write a FEFF input file. FEFF is a real-space Green's function code that employs a full multiple-scattering formalism \cite{FEFF9.6} for calculating the polarized VtC-XES. The following procedure is laid out in a diagram in Figure \ref{fig:Corvus-workflow}. Corvus starts by using Pymatgen \cite{Ong.etal2013} to predict the oxidation states of all the atoms in a unit cell, which are in turn used to construct the weighted quadrupole matrix as defined in eq \ref{eq:weighted_quad}. An important note is that for crystal structures that contain multiple instances of the same 3$d$ transition metal, the quadrupole tensors are averaged over all sites before diagonalizing. The orthogonal basis that comes from diagonalizing the quadrupole tensor is then used as the basis vectors for the subsequent FEFF VtC-XES calculation, and the generation of all subsequent features.

FEFF input files are programmatically generated for each individual cluster. The self-consistent field and full-multiple scattering radii used for the FEFF calculations are dynamically set to contain the closest 30 atoms to the central atom in each calculation, which helps achieve consistent convergence. All calculations were performed in the presence of a core hole, consistent with previous theoretical treatments of VtC-XES and the final-state rule \cite{mortensen-2017}. It should be noted that we also limit the number of unique potentials so that the absorbing atom in the center of the cluster is assigned one unique potential, while all other atoms have a potential which is only unique to their atomic number. The polarized output spectra are used to calculate the SAMS according to eqs \ref{eq:SAM} and \ref{eq:SAMS} to produce a target property that our model tries to predict. The total computation time across the full dataset took approximately 3300 hours of CPU time. 

% Figure \ref{fig:QAMS_Distribution} shows the distribution and Interquartile Range (IQR) of the QAMS for each transition metal. Figure \ref{fig:SAMS_Distribution} shows the distribution and IQR of the SAMS for each transition metal. We note that for both the QAMS and SAMS, scandium demonstrates the lowest mean and variance of any of the transition metal systems. 

% \begin{figure}
%     \centering
    
%     \begin{subfigure}{0.5\textwidth}
%     \includegraphics[width=1\linewidth]{figures/TM_distribution_bar.png} 
%     \caption{Distribution of the 3$d$ transition metals in the dataset, showing the total number of structures containing each metal.}
%     \label{fig:TM_distribution}
%     \end{subfigure}
    
%     \begin{subfigure}{0.5\textwidth}
%     \includegraphics[width=1\linewidth]{figures/QAMS_violin_plot.png} 
%     \caption{Distribution of QAMS grouped by transition metal}
%     \label{fig:QAMS_Distribution}
%     \end{subfigure}

%     \begin{subfigure}{0.5\textwidth}
%     \includegraphics[width=1\linewidth]{figures/SAMS_violin_plot.png} 
%     \caption{Distribution of SAMS grouped by transition metal}
%     \label{fig:SAMS_Distribution}
%     \end{subfigure}

% \caption{}
% \label{fig:TM_dataset_distribution}
% \end{figure}

\subsection{Machine Learning}

In this work, we employ a random forest regressor (RFR) \cite{breiman2001random} to predict the SAMS of 3$d$ transition metal clusters. An RFR is an ensemble learning method in which many decision trees are trained on bootstrapped samples of the data \cite{glassner-2021}, and the final prediction is obtained as the average of their outputs. Each tree is built through recursive binary splitting of the feature space, where only a randomly selected subset of features is considered at each split, thereby reducing correlation among trees and improving model robustness \cite{james2023isl, hastie2009elements, loh2014fifty}. This model architecture provides a practical compromise between accuracy, interpretability, and computational cost, making it well suited for relating the constructed features to the anisotropy observed in VtC-XES. The RFR implementation was constructed using the scikit-learn python library \cite{sklearn-rfr}. 

\subsection{Model Training and Hyperparameter Optimization}

% \begin{figure}
%     \centering
%     \includegraphics[width=\linewidth]{figures/sams_distribution.png}
%     \caption{Distribution of the target values for the training set ($y_{\mathrm{train}}$) and the test set ($y_{\mathrm{test}}$).}
%     \label{fig:sams_test_train_distribution}
% \end{figure}

The machine learning pipeline begins with the dataset being partitioned into training and testing subsets, with $80\%$ of the data used for training and the remaining $20\%$ reserved for testing. Hyperparameters were determined using a randomized grid search with $k$-fold cross-validation. Key RFR hyperparameters include the number and depth of trees, minimum samples per split/leaf, and the number of features considered at each split. We use bootstrap sampling (with optional subsampling) and squared error as the splitting criterion. In defining the hyperparameter space $\mathcal{G}$, we constrain the RFR hyperparameter space (e.g., depth, number of trees, and minimum samples per split/leaf) to mitigate overfitting and avoid memorization of training data. The constraints are defined in Table \ref{tab:hyperparam_constraints}.

\begin{table}[htbp]
\centering
\caption{Hyperparameter search constraints used during optimization of the RFR model.}
\label{tab:hyperparam_constraints}
\begin{tabular}{ll}
\toprule
\textbf{Hyperparameter} & \textbf{Constraint (Search Space)} \\
\midrule
\texttt{bootstrap}            & \texttt{[True]} \\
\texttt{criterion}            & \texttt{Fixed: squared\_error} \\
\texttt{max\_depth}           & \texttt{\{6, 8, 10, 12\}} \\
\texttt{max\_features}        & \texttt{\{"sqrt", "log2", 0.3\}} \\
\texttt{max\_samples}         & \texttt{Uniform(0.6, 0.8)} \\
\texttt{min\_samples\_leaf}   & \texttt{randint(5, 25)} \\
\texttt{min\_samples\_split}  & \texttt{randint(10, 50)} \\
\texttt{n\_estimators}        & \texttt{randint(300, 700)} \\
\bottomrule
\end{tabular}
\end{table}

We also employ $k$-fold cross-validation to further mitigate overfitting. The final performance estimate is obtained by averaging the results across all $k$ folds. By repeatedly shuffling and partitioning the data in this manner, we obtain a more reliable estimate of the model’s generalization performance on unseen data. For 100 randomly selected hyperparameter configuration from $\mathcal{G}$, performance was evaluated using $k$-fold cross-validation with the coefficient of determination ($R^2$) as the scoring metric. The highest performing hyperparameters were chosen for our model and are listed in Table \ref{tab:fixed_hyperparams}.

% We also employ $k$-fold cross-validation to further mitigate overfitting. This technique divides the training dataset into $k$ subsets, using $k-1$ subsets for training and the remaining subset for testing. The model is trained $k$ times, each time with a different subset serving as the test set, and a performance metric is computed in each iteration. The final performance estimate is obtained by averaging the results across all $k$ folds. By repeatedly shuffling and partitioning the data in this manner, we obtain a more reliable estimate of the model’s generalization performance on unseen data. From the hyperparameter space $\mathcal{G}$, we uniformly sampled up to 100 candidate configurations. For each configuration $\theta \in \mathcal{G}$, performance was evaluated using $k$-fold cross-validation with the coefficient of determination ($R^2$) as the scoring metric. The hyperparameters which performed the best are listed in Table \ref{tab:fixed_hyperparams}.

\begin{table}[htbp]
\centering
\caption{Fixed hyperparameters used for the RFR model.}
\label{tab:fixed_hyperparams}
\begin{tabular}{ll}
\toprule
\textbf{Hyperparameter} & \textbf{Value} \\
\midrule
\texttt{bootstrap}          & \texttt{True} \\
\texttt{criterion}          & \texttt{squared\_error} \\
\texttt{max\_depth}         & \texttt{12} \\
\texttt{max\_features}      & \texttt{0.3} \\
\texttt{max\_samples}       & \texttt{0.6676} \\
\texttt{min\_samples\_leaf} & \texttt{6} \\
\texttt{min\_samples\_split} & \texttt{11} \\
\texttt{n\_estimators}      & \texttt{587} \\
\bottomrule
\end{tabular}
\end{table}

\section{Results \& Discussion}
\label{sec:RandD}

\subsection{Model Results}

We use the coefficient of determination $R^2$ \cite{chicco2021rsq} to evaluate model performance along with the mean absolute error (MAE). The $R^2$ metric quantifies the fraction of variance in the target variable explained by the model predictions, while the MAE provides an alternative performance metric through the average absolute prediction error.

\begin{table}[H]
    \centering
    \caption{Performance metrics ($R^2$ and MAE) for the RFR model on the training and testing sets, with $\Delta$ (test -- train).}
    \label{tab:rfr_metrics}
    \begin{tabular}{lccc}
        \toprule
        \textbf{Metric} & \textbf{Training} & \textbf{Testing} & \textbf{$\Delta$} \\
        \midrule
        $R^2$ & 0.8526 & 0.7935 & $-0.0591$ \\
        MAE   & 0.0256 & 0.0298 & $+0.0042$ \\
        \bottomrule
    \end{tabular}
\end{table}

\begin{figure}[H]
    \centering
    \includegraphics[width=0.5\textwidth]{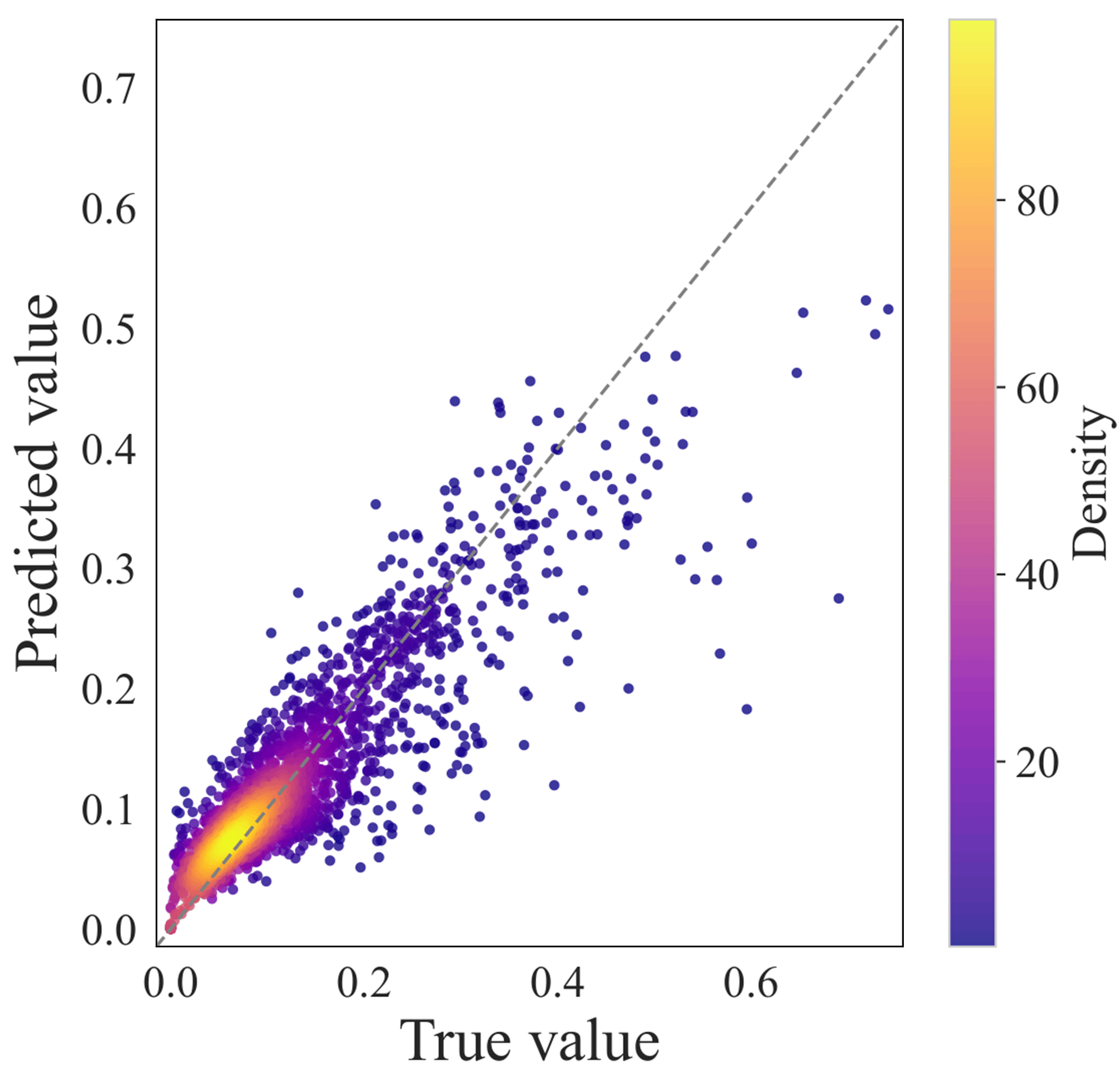}
    \caption{Predicted vs. True SAMS values from the trained RFR model acting on the testing set. The dashed diagonal line represents the ideal case where predictions exactly match the actual values, and the color scale indicates point density.}
    \label{fig:model_results}
\end{figure}

As shown in Table \ref{tab:rfr_metrics} our model achieves a high predictive accuracy, and the small difference between training and testing performance, $\Delta R^2 = -0.0591$ and $\Delta \mathrm{MAE} = +0.0042$, indicates that the model generalizes well to unseen data. The relatively high $R^2$ score ($0.79$ on the testing set) demonstrates that the model captures the majority of the variance in the SAMS from the chosen chemical and geometric input features. This level of predictive performance is consistent with other recent applications of machine learning in materials spectroscopy and property prediction, where $R^2$ values in the range of $0.7$–$0.9$ are typically reported for regression tasks involving complex, correlated physical quantities (e.g. chemical interpretation of near-edge x-ray absorption structure) \cite{torrisi-2020, iwayama-2022, guda-2021}.

Figure \ref{fig:model_results} shows the results of the trained model run on the test dataset. The vertical axis is the predicted SAMS value while the horizontal axis is the true SAMS value. We see that the distribution is strongly skewed along the diagonal, which is consistent with the high $R^2$ performance. Most of the materials are weakly anisotropic or not anisotropic at all, leading to a large density of points near the origin of the plot. Despite its strong overall performance, some systematic issues are evident. The model tends to slightly under predict SAMS values at the higher end of the distribution, while predictions for smaller SAMS values exhibit higher density and tighter clustering. This bias likely reflects the imbalance in the available data as shown in Figure \ref{fig:TM_dataset_distribution}, where extremely high-anisotropy configurations are less common. 

% Additionally, certain elements such as Sc and Ti possess narrower electronic structure variations across chemical environments, and may contribute disproportionately to the low-value region and thus influence the learned mapping.

\subsection{Model Interpretability}

RFR feature contributions can be assessed using variable importance measures \cite{strobl-2007}. Here, we focus on permutation feature importance (PFI), which avoids biases present in impurity-based measures, particularly for features with many unique values. This is relevant for our data, where SAMS values can take any real number between 0 and 1. Impurity-based measures also depend on training-set statistics and therefore may not reflect predictive performance on unseen data. PFI quantifies a feature’s importance by measuring the change in model performance after randomly permuting its values. A more detailed discussion of PFI can be found in \cite{molnar2025}. At a high level, PFI works by shuffling a feature’s values among all data points to break its relationship with the target and observing how much the model worsens. If performance degrades substantially, the feature is considered more important. Here the metric used to quantify the PFI is the negative mean squared error (NMSE), defined in Eq.~\ref{eq:nmse},

% PFI measures the importance of a feature by quantifying the change in model performance after the feature has been randomly permuted. In simple terms, this means shuffling the feature’s values among all data points to break its relationship with the target and observing how much the model worsens. If performance degrades substantially, the feature is considered more important. A more detailed discussion of PFI can be found in \cite{molnar2025}. In this work we use the negative mean squared error (NMSE), defined in Eq.~\ref{eq:nmse}:  

\begin{equation}
\text{NMSE}(y, \hat{y}) = -\frac{1}{n}\sum_{i=1}^{n} \big(y_i - \hat{y}_i\big)^2
\label{eq:nmse}
\end{equation}

where $y_i$ are the true values, $\hat{y}_i$ are the predicted values, and $n$ is the number of samples. We adopt NMSE following the convention in scikit-learn, where higher scores indicate better performance. Since the mean squared error is minimized, its negation ensures that values closer to zero represent better performance, while more negative values indicate worse performance~\cite{sklearn_permutation_importance}.  

Let $\text{NMSE}_0$ denote a baseline score of the model on the dataset. For each feature $j$, we construct a permuted dataset $X^{\pi(j)}$ by randomly shuffling the entries in column $j$ while leaving all other columns unchanged. We then recompute the score as shown in Eq.~\ref{eq:nmse_j}.  

\begin{equation}
\text{NMSE}_j = \text{NMSE}(y, \hat{y}^{\pi(j)})
\label{eq:nmse_j}
\end{equation}

The PFI for feature $j$ in a single permutation trial is defined in Eq.~\ref{eq:pfi_single}.  

\begin{equation}
\text{PFI}_j = \text{NMSE}_0 - \text{NMSE}_j
\label{eq:pfi_single}
\end{equation}

This process is repeated $R$ times with different random permutations, and the average importance is taken as in Eq.~\ref{eq:pfi_avg}.  

\begin{equation}
\overline{\text{PFI}}_j = \frac{1}{R}\sum_{r=1}^R \Big( \text{NMSE}_0 - \text{NMSE}_j^{(r)} \Big)
\label{eq:pfi_avg}
\end{equation}

\begin{figure}[htbp]
    \centering
    \includegraphics[width=0.47\textwidth]{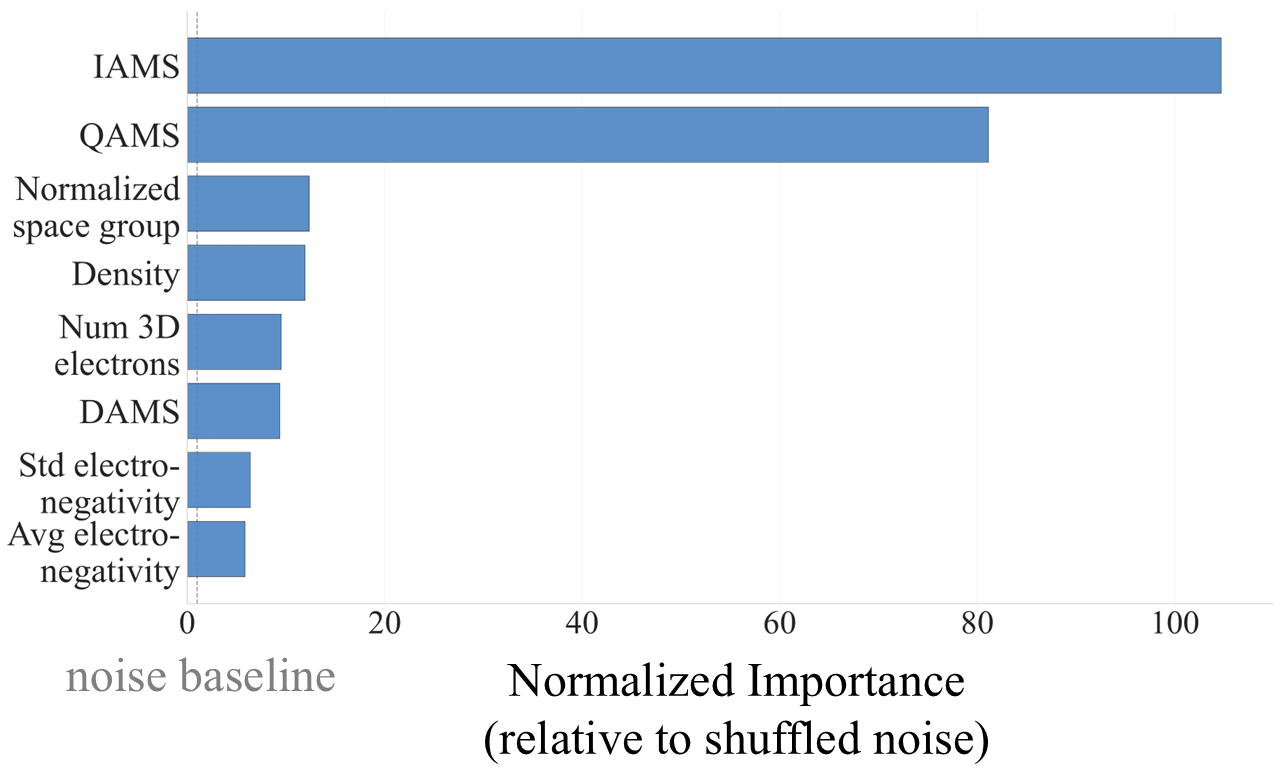}
    \caption{Noise-normalized permutation feature importance for the Random Forest Regressor. Bars represent the mean decrease in performance (increase in MSE) when each feature is permuted, normalized by the importance of a random noise baseline (set to 1). Values greater than one indicate predictive signal beyond random noise.}
    \label{fig:feature_importance}
\end{figure}

A feature ($j$) is considered important if the average importance, $\overline{\text{PFI}}_j$, is large, as this indicates a substantial degradation in performance when the feature is permuted~\cite{molnar2025interpretable}. However, the notion of “large” depends on the scale of the evaluation metric. Because the raw magnitudes of $\overline{\text{PFI}}_j$ depend on the dataset and model, they are not directly comparable across problems. To provide a meaningful scale, we normalize by the importance of a random noise feature, $\overline{\text{PFI}}_{\text{noise}}$, to provide a noise-normalized $\widetilde{\text{PFI}}_j$ in eq~\ref{eq:norm_PFI}.

\begin{equation}
    \widetilde{\text{PFI}}_j = \frac{\overline{\text{PFI}}_j}{\overline{\text{PFI}}_{\text{noise}}}
    \label{eq:norm_PFI}
\end{equation}

A $\widetilde{\text{PFI}}_j \approx 1$ implies that feature $j$ is no more predictive than random noise, while $\widetilde{\text{PFI}}_j > 1$ indicates predictive power beyond noise. For example, $\widetilde{\text{PFI}}_j = 5$ means that feature $j$ is five times as informative as the noise baseline. As shown in Figure \ref{fig:feature_importance}, the anisotropy-based descriptors IAMS and QAMS dominate the model’s predictive performance, consistent with their direct link to polarization-dependent spectral asymmetries discussed in Section \ref{sec:methods}. The modest importance of geometric and chemical descriptors such as the normalized space group, density, and DAMS indicates that the model captures secondary contributions from the structural environment. The slight advantage of IAMS over QAMS likely reflects reduced uncertainty in quantities derived directly from computed spectra compared to oxidation-state–dependent charge estimates. 

\subsection{Model Performance for 3$d$ Transition Metals}

\begin{figure}[t]
    \centering
    \includegraphics[width=\linewidth]{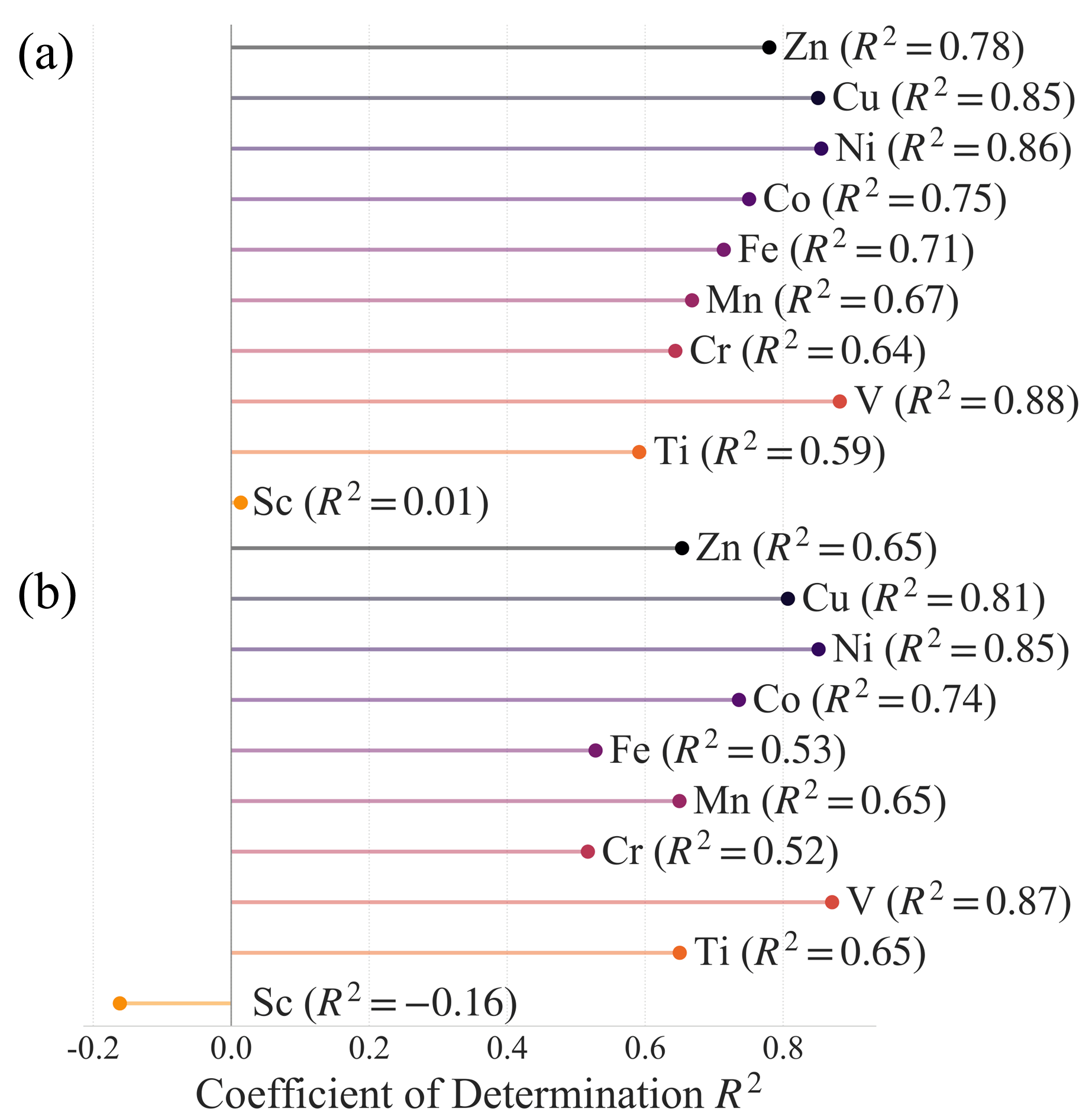}
    \caption{
    Per-element performance of the random forest regressor evaluated on the test set. Ordered from largest (top) to smallest (bottom) atomic number. (a) shows the coefficient of determination $R^2$ computed separately for each metal
    using the combined (unbalanced) training dataset, and (b) shows the corresponding
    $R^2$ values obtained using the balanced dataset.
    In both panels, the model is trained on all elements and evaluated on element-specific
    subsets of the test data.
    %The horizontal lines indicate the reference from $R^2 = 0$ to the achieved value for each element.
    }
    \label{fig:combined_balanced_indv_metal}
\end{figure}

\hspace{1em} To better understand the limitations of the model, we analyzed its performance on a per-element basis. Specifically, we computed the coefficient of determination, $R^2$, for each individual metal using subsets of the test data corresponding to a single element, while keeping the random forest regressor (RFR) trained on the full dataset. The resulting per-element $R^2$ values are shown in Figure \ref{fig:combined_balanced_indv_metal}a. The coefficient of determination is defined in eq \ref{eq:r2},

\begin{equation}
    R^2 = 1 - \frac{\sum_i \left(y_i - \hat{y}_i\right)^2}{\sum_i \left(y_i - \bar{y}\right)^2},
    \label{eq:r2}
\end{equation}

where $y_i$ are the true target values, $\hat{y}_i$ are the corresponding model predictions, and $\bar{y}$ is the mean of the target values for the subset under consideration.

\begin{figure}[!t]
    \centering
    \includegraphics[width=\linewidth]{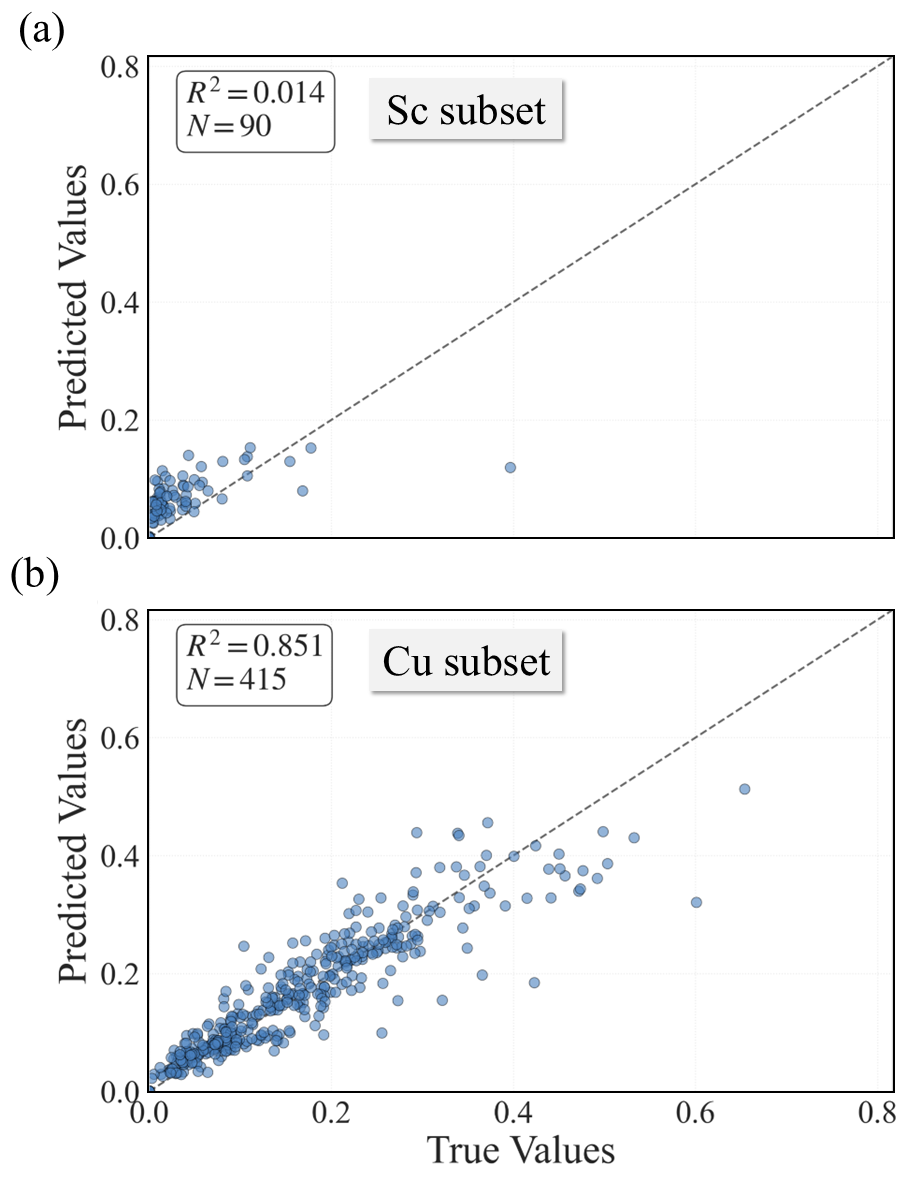}
    \caption{
    Predicted versus true SAMS values for element-specific subsets of the test data. (a) Shows the Sc subset, for which the model achieves a low coefficient of determination
    ($R^2 = 0.014$, $N = 90$), indicating performance only marginally better than a mean predictor. (b) Shows the Cu subset, where the model performs substantially better
    ($R^2 = 0.851$, $N = 415$), with predictions closely following the ideal $y = x$ relationship
    (dashed line).
    In both panels, the model is trained on the full multi-element dataset and evaluated on
    element-specific test subsets.
    }
\label{fig:Sc_Cu_combined_test_performance}
\end{figure}

Among all elements, Sc exhibits the poorest performance, with an $R^2$ value close to zero, indicating that the model provides little improvement over a naive mean predictor for this element. Figure \ref{fig:Sc_Cu_combined_test_performance} shows predicted versus true SAMS values for the Sc Figure \ref{fig:Sc_Cu_combined_test_performance}a and Cu Figure \ref{fig:Sc_Cu_combined_test_performance}b subsets of the test data. Showcasing the difference between a subset of the data that the model performs well on and one that it does not.

For Sc Figure \ref{fig:Sc_Cu_combined_test_performance}a, the coefficient of determination is $R^2 \approx 0.01$, which corresponds to a ratio of squared errors $\approx 0.99$. This indicates that the model reduces the error relative to a mean predictor by only about 1\%. The small but positive $R^2$ value suggests that the model predictions are slightly shifted relative to the Sc-specific mean, consistent with a bias toward the global mean of the full multi-element dataset. In contrast to Sc, Cu Figure \ref{fig:Sc_Cu_combined_test_performance}b exhibits excellent performance, with an $R^2$ of $0.851$, demonstrating that the model can accurately learn patterns when sufficient representative data is present.

Next, we investigate if the poor performance of Sc may be due to under representation of this element in the Materials Project database. To address this, we constructed a balanced data set with an equal number of compounds (444 compounds from each element, see Figure \ref{fig:TM_dataset_distribution}a). The training and testing sets were also split such that there was an even distribution of each element, with 354 samples in the training, and 90 samples in the test set. The results of this are presented in Figure~\ref{fig:combined_balanced_indv_metal}b, where the performance for Sc significantly worsens, suggesting that its poor predictive behavior likely stems from intrinsic differences in the characteristics of Sc compared to the other 3$d$ transition metals rather than data imbalance.

One possible explanation for the poorer performance on Sc is that its chemical behavior differs significantly from the dataset's majority. Unlike most 3$d$ transition metals, Sc exhibits nearly exclusive $3+$ oxidation, resulting in a $3d^0$ closed-shell configuration with weak ligand-field effects and minimal metal–ligand covalency. Consequently, its spectra are less anisotropic and structurally sensitive, differing fundamentally from the more covalent and partially filled $3d$ systems the model was trained on.

\section{Conclusion}
\label{sec:conclusion}

In this work we demonstrated that a random forest regressor model is able to accurately predict a continuous anisotropy parameter using structural and chemical information extracted from crystallographic structures. 
This result provides an important step in creating new opportunities for machine learning driven materials discovery, as well as the design of materials with targeted spectroscopic properties. The model generalizes well to any crystalline 3$d$ transition metal structure (with the exception of Sc), but could easily be adapted to treat molecular systems as well by replacing or removing the normed space group feature. 

It is also important to emphasize that the model should not be applied outside the scope in which it has been trained. While the approach demonstrates strong predictive capability within crystalline 3$d$ transition metal compounds, it is not applicable to materials with significantly different chemistry, dimensionality, or bonding environments (e.g., amorphous solids, $f$-electron oxides) without retraining or adaptation. In such cases, careful validation and retraining with an appropriately curated dataset are necessary before drawing conclusions.

Future work includes extending our machine learning model to predict anisotropy in other x-ray methods, such as XAS. In particular, many of the structural and chemical descriptors that drive anisotropy in XES are also expected to influence absorption edge features, suggesting that the framework developed here could be transferable across spectroscopy techniques. It may be possible to apply a transfer learning approach so that the already trained XES anisotropy model can be fine-tuned for predicting anisotropy in a different technique, which could dramatically reduce the computational cost compared to training a new model from scratch.

Another important step is expanding the scope of the training data itself. Incorporating systems with controlled defects, surface terminations, or varying dimensionality would allow the model to capture a wider range of realistic environments, improving its applicability to experimental conditions. Beyond methodological development, coupling the model with high-throughput structure generation and screening pipelines could open new opportunities for accelerated materials discovery. For example, one could rapidly explore chemical substitutions, strain effects, or heterostructure geometries and identify candidates with large or tunable anisotropy.

%Lowering both the computational overhead of spectral calculation, and the amount of time it takes for researchers to find symmetrically related systems, will help accelerate the discovery of exotic materials and their unexplored applications.

%Additionally, the generation of the input features is robust enough to be able to handle defects, such as vacancies or dislocations

% Our dataset of approximately 10000 experimentally observed materials achieved a RMSE of (\textbf{NUMBER}), and MAE of (\textbf{NUMBER}) when using the DAMS, QAMS, IAMS, among other factors, to demonstrate the anisotropy of a crystal system using linearly polarized VtC-XES. 

%The results from this work will allow researchers from other fields interested in materials properties to choose from a wide selection of 3$d$ transition metal systems to best tailor their experiments. 

\section{Acknowledgments}

JJK and JJR acknowledge support from the Theory Center for Materials and Energy Spectroscopies (TIMES) at SLAC funded by DOE BES Contract DE-AC02-76SF00515. CAC was supported by the National Science Foundation Graduate Research Fellowship Program under Grant No. DGE-2140004. GTS was supported by funding from the U.S. Department of Energy in the Nuclear Energy University Program under Contract No. DE-NE0009158. Any opinions, findings, and conclusions or recommendations expressed in this material are those of the author(s) and do not necessarily reflect the views of the National Science Foundation or the U.S. Department of Energy. This work was partly facilitated through the use of advanced computational infrastructure provided by the Hyak supercomputer system at the University of Washington.

\newpage

% \subsection{FEFF used for calculation of polarized VtC-XES}

% Reasonable qualitative and quantitative fitting of theoretical spectra to experiment must be demonstrated with machine learning methods to ensure that large datasets can provide physically reliable results. Here we show two polarized VtC-XES experimental results from the literature, and compare our proposed theory calculations from our workflow. 

\newpage

\bibliography{report}

@article{glatzel-2004,
	author = {Glatzel, Pieter and Bergmann, Uwe},
	journal = {Coordination Chemistry Reviews},
	month = {6},
	number = {1-2},
	pages = {65--95},
	title = {{High resolution 1s core hole X-ray spectroscopy in 3d transition metal complexes—electronic and structural information}},
	volume = {249},
	year = {2004},
	doi = {10.1016/j.ccr.2004.04.011},
	url = {https://doi.org/10.1016/j.ccr.2004.04.011},
}

@article{mortensen-2017,
	author = {Mortensen, D. R. and Seidler, G. T. and Kas, Joshua J. and Govind, Niranjan and Schwartz, Craig P. and Pemmaraju, Sri and Prendergast, David G.},
	journal = {Phys. Rev. B},
	month = {9},
	number = {12},
	title = {{Benchmark results and theoretical treatments for valence-to-core x-ray emission spectroscopy in transition metal compounds}},
	volume = {96},
	year = {2017},
	doi = {10.1103/physrevb.96.125136},
	url = {https://doi.org/10.1103/physrevb.96.125136},
}

@manual{FEFF9.6,
  title        = {FEFF9.6 User’s Guide},
  author       = {{The FEFF Project}},
  organization = {Department of Physics, University of Washington},
  year         = {2013},
  month        = {February},
  note         = {feff version 9.6.4, updated February 2, 2013}
}

@article{pollock-2015,
	author = {Pollock, Christopher J. and DeBeer, Serena},
	journal = {Accounts of Chemical Research},
	month = {9},
	number = {11},
	pages = {2967--2975},
	title = {{Insights into the Geometric and Electronic Structure of Transition Metal Centers from Valence-to-Core X-ray Emission Spectroscopy}},
	volume = {48},
	year = {2015},
	doi = {10.1021/acs.accounts.5b00309},
	url = {https://doi.org/10.1021/acs.accounts.5b00309},
}

@article{van-der-laan-2014,
	author = {Van Der Laan, Gerrit and Figueroa, Adriana I.},
	journal = {Coordination Chemistry Reviews},
	month = {3},
	pages = {95--129},
	title = {{X-ray magnetic circular dichroism - A versatile tool to study magnetism}},
	volume = {277-278},
	year = {2014},
	doi = {10.1016/j.ccr.2014.03.018},
	url = {https://doi.org/10.1016/j.ccr.2014.03.018},
}

@book{stohr-1992,
	author = {Stöhr, Joachim},
	month = {1},
	title = {{NEXAFS Spectroscopy}},
	year = {1992},
    publisher = {Springer Series in Surface Sciences},
	doi = {10.1007/978-3-662-02853-7},
	url = {https://doi.org/10.1007/978-3-662-02853-7},
}

@article{Kas.etal2021a,
  title = {Advanced Calculations of {{X-ray}} Spectroscopies with {{{\emph{FEFF10}}}} and {{Corvus}}},
  author = {Kas, J. J. and Vila, F. D. and Pemmaraju, C. D. and Tan, T. S. and Rehr, J. J.},
  date = {2021-11-01},
  journal = {Journal of Synchrotron Radiation},
  volume = {28},
  number = {6},
  pages = {1801--1810},
  issn = {1600-5775},
  year = {2021},
  month = {11},
  doi = {10.1107/S1600577521008614},
}

@article{timoshenko-2018,
	author = {Timoshenko, Janis and Anspoks, Andris and Cintins, Arturs and Kuzmin, Alexei and Purans, Juris and Frenkel, Anatoly I.},
	journal = {Physical Review Letters},
    volume = {120},
    issue = {22},
    pages = {225502},
    numpages = {6},
    year = {2018},
    month = {May},
	title = {{Neural network approach for characterizing structural transformations by X-Ray absorption fine structure spectroscopy}},
	doi = {10.1103/physrevlett.120.225502},
	url = {https://doi.org/10.1103/physrevlett.120.225502},
}

@article{rankine-2020,
	author = {Rankine, C. D. and Madkhali, M. M. M. and Penfold, T. J.},
	journal = {The Journal of Physical Chemistry A},
	month = {5},
	number = {21},
	pages = {4263--4270},
	title = {{A deep neural network for the rapid prediction of x-ray absorption spectra}},
	volume = {124},
	year = {2020},
	doi = {10.1021/acs.jpca.0c03723},
	url = {https://doi.org/10.1021/acs.jpca.0c03723},
}

@article{gastegger-2017,
	author = {Gastegger, Michael and Behler, Jörg and Marquetand, Philipp},
	journal = {Chemical Science},
	month = {1},
	number = {10},
	pages = {6924--6935},
	title = {{Machine learning molecular dynamics for the simulation of infrared spectra}},
	volume = {8},
	year = {2017},
	doi = {10.1039/c7sc02267k},
	url = {https://doi.org/10.1039/c7sc02267k},
}

@article{lou-2024,
	author = {Lou, Yuchen and Ganose, Alex M.},
	journal = {Faraday Discussions},
	month = {7},
    volume = {256},
    pages = {255-274},
	title = {{Discovery of highly anisotropic dielectric crystals with equivariant graph neural networks}},
	year = {2024},
	doi = {10.1039/d4fd00096j},
	url = {https://doi.org/10.1039/d4fd00096j},
}

@article{wen-2024,
	author = {Wen, Mingjian and Horton, Matthew K. and Munro, Jason M. and Huck, Patrick and Persson, Kristin A.},
	journal = {Digital Discovery},
	month = {1},
	number = {5},
	pages = {869--882},
	title = {{An equivariant graph neural network for the elasticity tensors of all seven crystal systems}},
	volume = {3},
	year = {2024},
	doi = {10.1039/d3dd00233k},
	url = {https://doi.org/10.1039/d3dd00233k},
}

@article{Ong.etal2013,
  title = {Python {{Materials Genomics}} (Pymatgen): {{A}} Robust, Open-Source Python Library for Materials Analysis},
  shorttitle = {Python {{Materials Genomics}} (Pymatgen)},
  author = {Ong, Shyue Ping and Richards, William Davidson and Jain, Anubhav and Hautier, Geoffroy and Kocher, Michael and Cholia, Shreyas and Gunter, Dan and Chevrier, Vincent L. and Persson, Kristin A. and Ceder, Gerbrand},
  date = {2013-02-01},
  year = {2013},
  journal = {Computational Materials Science},
  volume = {68},
  pages = {314--319},
  issn = {0927-0256},
  doi = {10.1016/j.commatsci.2012.10.028},
}

@article{bishop-2007,
	author = {Bishop, Christopher M.},
	journal = {Journal of Electronic Imaging},
	month = {1},
	number = {4},
	pages = {049901},
	title = {{Pattern recognition and machine learning}},
	volume = {16},
	year = {2007},
	doi = {10.1117/1.2819119},
	url = {https://doi.org/10.1117/1.2819119},
}

@article{pithan-2023,
	author = {Pithan, Linus and Starostin, Vladimir and Mareček, David and Petersdorf, Lukas and Völter, Constantin and Munteanu, Valentin and Jankowski, Maciej and Konovalov, Oleg and Gerlach, Alexander and Hinderhofer, Alexander and Murphy, Bridget and Kowarik, Stefan and Schreiber, Frank},
	journal = {Journal of Synchrotron Radiation},
	month = {9},
	number = {6},
	pages = {1064--1075},
	title = {{Closing the loop: autonomous experiments enabled by machine-learning-based online data analysis in synchrotron beamline environments}},
	volume = {30},
	year = {2023},
	doi = {10.1107/s160057752300749x},
	url = {https://doi.org/10.1107/s160057752300749x},
}

@article{Zheng.etal2020,
  title = {Random {{Forest Models}} for {{Accurate Identification}} of {{Coordination Environments}} from {{X-Ray Absorption Near-Edge Structure}}},
  author = {Zheng, Chen and Chen, Chi and Chen, Yiming and Ong, Shyue Ping},
  date = {2020-05},
  journal = {Patterns},
  year = {2020},
  month = {May},
  volume = {1},
  number = {2},
  pages = {100013},
  issn = {26663899},
  doi = {10.1016/j.patter.2020.100013},
  url = {https://linkinghub.elsevier.com/retrieve/pii/S2666389920300131},
  urldate = {2024-11-01},
  langid = {english},
}

@article{sendek-2018,
	author = {Sendek, Austin D. and Cubuk, Ekin D. and Antoniuk, Evan R. and Cheon, Gowoon and Cui, Yi and Reed, Evan J.},
	journal = {Chemistry of Materials},
	month = {11},
	number = {2},
	pages = {342--352},
	title = {{Machine Learning-Assisted discovery of solid Li-Ion conducting materials}},
	volume = {31},
	year = {2018},
	doi = {10.1021/acs.chemmater.8b03272},
	url = {https://doi.org/10.1021/acs.chemmater.8b03272},
}

@article{shen-2022,
	author = {Shen, Lei and Zhou, Jun and Yang, Tong and Yang, Ming and Feng, Yuan Ping},
	journal = {Accounts of Materials Research},
	month = {5},
	number = {6},
	pages = {572--583},
	title = {{High-Throughput computational discovery and intelligent design of Two-Dimensional functional materials for various applications}},
	volume = {3},
	year = {2022},
	doi = {10.1021/accountsmr.1c00246},
	url = {https://doi.org/10.1021/accountsmr.1c00246},
}

@article{hirohata-2025,
	author = {Hirohata, Atsufumi and Koizumi, Hiroki and Roy, Tufan and Tsujikawa, Masahito and Mizukami, Shigemi and Nawa, Kenji and Shirai, Masafumi},
	journal = {NPJ Spintronics},
	month = {7},
	number = {1},
	title = {{Machine learning for the development of new materials for a magnetic tunnel junction}},
	volume = {3},
	year = {2025},
	doi = {10.1038/s44306-025-00094-z},
	url = {https://www.nature.com/articles/s44306-025-00094-z},
}

@article{bergmann-2002,
	author = {Bergmann, U. and Bendix, J. and Glatzel, P. and Gray, H. B. and Cramer, S. P.},
	journal = {The Journal of Chemical Physics},
	month = {2},
	number = {5},
	pages = {2011-2015},
	title = {{Anisotropic valence→core x-ray fluorescence from a [Rh(en)3][Mn(N)(CN)5]H2O single crystal: Experimental results and density functional calculations}},
	volume = {116},
	year = {2002},
	doi = {10.1063/1.1419062},
	url = {https://doi.org/10.1063/1.1419062},
}

@article{jansing-2016,
	author = {Jansing, C. and Mertins, H.-Ch. and Gilbert, M. and Wahab, H. and Timmers, H. and Choi, S.-h. and Gaupp, A. and Krivenkov, M. and Varykhalov, A. and Rader, O. and Legut, D. and Oppeneer, P. M.},
	journal = {Phys. Rev. B},
	month = {7},
	number = {4},
	title = {{X-ray natural birefringence in reflection from graphene}},
	volume = {94},
	year = {2016},
	doi = {10.1103/physrevb.94.045422},
	url = {https://doi.org/10.1103/physrevb.94.045422},
}

@article{van-der-laan-2011,
	author = {Van Der Laan, G. and Telling, N. D. and Potenza, A. and Dhesi, S. S. and Arenholz, E.},
	journal = {Physical Review B},
	month = {2},
	number = {6},
	title = {{Anisotropic x-ray magnetic linear dichroism and spectromicroscopy of interfacial Co/NiO(001)}},
	volume = {83},
	year = {2011},
	doi = {10.1103/physrevb.83.064409},
	url = {https://doi.org/10.1103/physrevb.83.064409},
}

@article{ketenoglu-2022,
	author = {Ketenoglu, Didem},
	journal = {X-Ray Spectrometry},
	month = {6},
	number = {5-6},
	pages = {422--443},
	title = {{A general overview and comparative interpretation on element‐specific X‐ray spectroscopy techniques: XPS, XAS, and XRS}},
	volume = {51},
	year = {2022},
	doi = {10.1002/xrs.3299},
	url = {https://doi.org/10.1002/xrs.3299},
}

@article{stohr-1999,
	author = {Stöhr, J.},
	journal = {Journal of Magnetism and Magnetic Materials},
	month = {10},
	number = {1-3},
	pages = {470--497},
	title = {{Exploring the microscopic origin of magnetic anisotropies with X-ray magnetic circular dichroism (XMCD) spectroscopy}},
	volume = {200},
	year = {1999},
	doi = {10.1016/s0304-8853(99)00407-2},
	url = {https://doi.org/10.1016/s0304-8853(99)00407-2},
}

@article{bianconi-1988,
	author = {Bianconi, A. and De Santis, M. and Di Cicco, A. and Flank, A. M. and Fontaine, A. and Lagarde, P. and Katayama-Yoshida, H. and Kotani, A. and Marcelli, A.},
	journal = {Physical review. B, Condensed matter},
	month = {10},
	number = {10},
	pages = {7196--7199},
	title = {{Symmetry of the 3d9 ligand hole induced by doping in YBa$_2$}},
	volume = {38},
	year = {1988},
	doi = {10.1103/physrevb.38.7196},
	url = {https://doi.org/10.1103/physrevb.38.7196},
}

@article{jo-2004,
	author = {Jo, Takeo},
	journal = {Journal of Electron Spectroscopy and Related Phenomena},
	month = {4},
	number = {1-2},
	pages = {99--106},
	title = {{Orbital states and polarized X-ray absorption}},
	volume = {136},
	year = {2004},
	doi = {10.1016/j.elspec.2004.02.136},
	url = {https://doi.org/10.1016/j.elspec.2004.02.136},
}

@article{breiman2001random,
  title={Random forests},
  author={Breiman, Leo},
  journal={Machine Learning},
  volume={45},
  number={1},
  pages={5--32},
  year={2001},
  publisher={Springer}
}

@article{loh2014fifty,
  title={Fifty years of classification and regression trees},
  author={Loh, Wei-Yin},
  journal={International Statistical Review},
  volume={82},
  number={3},
  pages={329--348},
  year={2014},
  publisher={Wiley Online Library}
}

@book{hastie2009elements,
  title={The Elements of Statistical Learning: Data Mining, Inference, and Prediction},
  author={Hastie, Trevor and Tibshirani, Robert and Friedman, Jerome},
  year={2009},
  publisher={Springer}
}

@article{trippe-2014,
	author = {Trippe, S.},
	journal = {arXiv (Cornell University)},
	month = {1},
	title = {{Polarization and Polarimetry: A review}},
	year = {2014},
	doi = {10.48550/arxiv.1401.1911},
	url = {https://arxiv.org/abs/1401.1911},
}

@article{fumagalli-2019,
	author = {Fumagalli, R. and Braicovich, L. and Minola, M. and Peng, Y. Y. and Kummer, K. and Betto, D. and Rossi, M. and Lefrançois, E. and Morawe, C. and Salluzzo, M. and Suzuki, H. and Yakhou, F. and Tacon, M. Le and Keimer, B. and Brookes, N. B. and Sala, M. Moretti and Ghiringhelli, G.},
	journal = {Phys. Rev. B},
	month = {4},
	number = {13},
	title = {{Polarization-resolved Cu L3 -edge resonant inelastic x-ray scattering of orbital and spin excitations in NdBa$_2$Cu$_3$O$_7$}},
	volume = {99},
	year = {2019},
	doi = {10.1103/physrevb.99.134517},
	url = {https://doi.org/10.1103/physrevb.99.134517},
}

@book{kim-2017,
	author = {Kim, Kwang-Je and Huang, Zhirong and Lindberg, Ryan},
	month = {3},
	title = {{Synchrotron radiation and Free-Electron lasers}},
	year = {2017},
    publisher = {Cambridge University Press},
	doi = {10.1017/9781316677377},
	url = {https://doi.org/10.1017/9781316677377},
}

@book{james2023isl,
  title     = {An Introduction to Statistical Learning: With Applications in Python},
  author    = {James, Gareth and Witten, Daniela and Hastie, Trevor and Tibshirani, Robert and Taylor, Jonathan},
  publisher = {Springer},
  year      = {2023},
  edition   = {2nd},
  isbn      = {978-1-0716-8759-5},
  note      = {See Section~8.2, p.~343}
}

@article{nascimento-2022,
	author = {Nascimento, Daniel R. and Govind, Niranjan},
	journal = {Physical Chemistry Chemical Physics},
	month = {1},
	number = {24},
	pages = {14680--14691},
	title = {{Computational approaches for XANES, VtC-XES, and RIXS using linear-response time-dependent density functional theory based methods}},
	volume = {24},
	year = {2022},
	doi = {10.1039/d2cp01132h},
	url = {https://pubs.rsc.org/en/content/articlelanding/2022/cp/d2cp01132h},
}

@article{roemelt-2024,
	author = {Roemelt, Christina and Peredkov, Sergey and Neese, Frank and Roemelt, Michael and DeBeer, Serena},
	journal = {Physical Chemistry Chemical Physics},
	month = {1},
	number = {29},
	pages = {19960--19975},
	title = {{Valence-to-Core X-Ray emission Spectroscopy of transition Metal tetrahalides: Mechanisms governing intensities}},
	volume = {26},
	year = {2024},
	doi = {10.1039/d4cp00967c},
	url = {https://pubs.rsc.org/en/content/articlelanding/2024/cp/d4cp00967c},
}

@article{chicco2021rsq,
  title        = {The coefficient of determination R-squared is more informative than SMAPE, MAE, MAPE, MSE and RMSE in regression analysis evaluation},
  author       = {Chicco, Davide and Warrens, Matthijs J and Jurman, Giuseppe},
  journal      = {PeerJ Computer Science},
  volume       = {7},
  pages        = {e623},
  year         = {2021},
  publisher    = {PeerJ Inc.},
  doi          = {10.7717/peerj-cs.623},
  pmid         = {34307865},
  pmcid        = {PMC8279135}
}

@misc{sklearn-rfr,
  author       = {{scikit-learn developers}},
  title        = {Random Forest Regressor},
  howpublished = {\textit{scikit-learn} 1.7.1 Documentation},
  year         = {2025},
  url          = {https://scikit-learn.org/stable/modules/generated/sklearn.ensemble.RandomForestRegressor.html},
  note         = {accessed August 13, 2025}
}

@article{geoghegan-2022,
	author = {Geoghegan, Blaise L. and Liu, Yang and Peredkov, Sergey and Dechert, Sebastian and Meyer, Franc and DeBeer, Serena and Cutsail, George E.},
	journal = {Journal of the American Chemical Society},
	month = {1},
	number = {6},
	pages = {2520--2534},
	title = {{Combining Valence-to-Core X-ray emission and Cu K-Edge X-ray absorption spectroscopies to experimentally assess oxidation state in organometallic Cu(I)/(II)/(III) complexes}},
	volume = {144},
	year = {2022},
	doi = {10.1021/jacs.1c09505},
	url = {https://doi.org/10.1021/jacs.1c09505},
}

@online{sklearn_permutation_importance,
  title={Permutation Importance vs Random Forest Feature Importance (MDI)},
  author={{Scikit-learn}},
  url={https://scikit-learn.org/stable/auto_examples/inspection/plot_permutation_importance.html},
  note={Accessed August 16, 2025}
}

@book{molnar2025interpretable,
  title={Interpretable Machine Learning: A Guide for Making Black Box Models Explainable},
  author={Molnar, Christoph},
  year={2025},
  edition={3rd},
  publisher={Leanpub},
  url={https://christophm.github.io/interpretable-ml-book/},
}

@book{sakurai1967advanced,
  title = {Advanced Quantum Mechanics},
  author = {Sakurai, J.J.},
  date = {1967},
  publisher = {Addison-Wesley},
  isbn = {978-81-7758-916-0},
  year={1967}
}

@article{Drager.Brummer1984,
  title = {Polarized {{X-Ray Emission Spectra}} of {{Single Crystals}}},
  author = {Dräger, G. and Brümmer, O.},
  date = {1984},
  journaltitle = {physica status solidi (b)},
  volume = {124},
  number = {1},
  pages = {11--28},
  issn = {1521-3951},
  doi = {10.1002/pssb.2221240102},
  url = {https://onlinelibrary.wiley.com/doi/abs/10.1002/pssb.2221240102},
  urldate = {2025-08-19},
  langid = {english}
}

@article{nielsen-2024,
	author = {Nielsen, Villads R. M. and Guennic, Boris Le and Sørensen, Thomas Just},
	journal = {The Journal of Physical Chemistry A},
	month = {6},
	number = {28},
	pages = {5740--5751},
	title = {{Evaluation of Point Group Symmetry in Lanthanide(III) Complexes: A New Implementation of a Continuous Symmetry Operation Measure with Autonomous Assignment of the Principal Axis}},
	volume = {128},
	year = {2024},
	doi = {10.1021/acs.jpca.4c00801},
	url = {https://doi.org/10.1021/acs.jpca.4c00801},
}

@article{brener-2024,
  title = {Anisotropic excitonic magnetism from discrete ${C}_{4}$ symmetry in ${\mathrm{CeRhIn}}_{5}$},
  author = {Brener, D. J. and Mallo, I. Rodriguez and Lane, H. and Rodriguez-Rivera, J. A. and Schmalzl, K. and Songvilay, M. and Guratinder, K. and Petrovic, C. and Stock, C.},
  journal = {Phys. Rev. B},
  volume = {110},
  issue = {6},
  pages = {064434},
  numpages = {15},
  year = {2024},
  month = {Aug},
  publisher = {American Physical Society},
  doi = {10.1103/PhysRevB.110.064434},
  url = {https://link.aps.org/doi/10.1103/PhysRevB.110.064434}
}

@book{watanabe-1966,
	author = {Watanabe, Hiroshi},
	title = {{Operator methods in ligand field theory}},
	year = {1966},
    publisher = {Prentice-Hall},
	url = {http://ci.nii.ac.jp/ncid/BA13037652},
}

@article{lafuerza-2020,
	author = {Lafuerza, Sara and Carlantuono, Andrea and Retegan, Marius and Glatzel, Pieter},
	journal = {Inorganic Chemistry},
	month = {8},
	number = {17},
	pages = {12518--12535},
	title = {{Chemical Sensitivity of K$\beta$ and K$\alpha$ X-ray Emission from a Systematic Investigation of Iron Compounds}},
	volume = {59},
	year = {2020},
	doi = {10.1021/acs.inorgchem.0c01620},
	url = {https://doi.org/10.1021/acs.inorgchem.0c01620},
}

@article{zutic-2004,
	author = {Žutić, Igor and Fabian, Jaroslav and Sarma, S. Das},
	journal = {Reviews of Modern Physics},
	month = {4},
	number = {2},
	pages = {323--410},
	title = {{Spintronics: Fundamentals and applications}},
	volume = {76},
	year = {2004},
	doi = {10.1103/revmodphys.76.323},
	url = {https://doi.org/10.1103/revmodphys.76.323},
}

@article{wang-2023,
	author = {Wang, Youwen and Luo, Nannan and Zeng, Jiang and Tang, Li-Ming and Chen, Ke-Qiu},
	journal = {Phys. Rev. B},
	month = {8},
	number = {5},
	title = {{Magnetic anisotropy and electric field induced magnetic phase transition in the van der Waals antiferromagnet CrSBr}},
	volume = {108},
	year = {2023},
	doi = {10.1103/physrevb.108.054401},
	url = {https://doi.org/10.1103/physrevb.108.054401},
}

@article{tombros-2008,
	author = {Tombros, N. and Tanabe, S. and Veligura, A. and Jozsa, C. and Popinciuc, M. and Jonkman, H. T. and Van Wees, B. J.},
	journal = {Physical Review Letters},
	month = {7},
	number = {4},
	title = {{Anisotropic spin relaxation in graphene}},
	volume = {101},
	year = {2008},
	doi = {10.1103/physrevlett.101.046601},
	url = {https://doi.org/10.1103/physrevlett.101.046601},
}

@article{Jain2013,
author = {Jain, Anubhav and Ong, Shyue Ping and Hautier, Geoffroy and Chen, Wei and Richards, William Davidson and Dacek, Stephen and Cholia, Shreyas and Gunter, Dan and Skinner, David and Ceder, Gerbrand and Persson, Kristin a.},
doi = {10.1063/1.4812323},
issn = {2166532X},
journal = {APL Materials},
number = {1},
pages = {011002},
title = {{The Materials Project: A materials genome approach to accelerating materials innovation}},
url = {http://link.aip.org/link/AMPADS/v1/i1/p011002/s1\&Agg=doi},
volume = {1},
year = {2013}
}

@article{gong_2017,
author = {Gong, Chuanhui and Zhang, Yuxi and Chen, Wei and Chu, Junwei and Lei, Tianyu and Pu, Junru and Dai, Liping and Wu, Chunyang and Cheng, Yuhua and Zhai, Tianyou and Li, Liang and Xiong, Jie},
title = {Electronic and Optoelectronic Applications Based on 2D Novel Anisotropic Transition Metal Dichalcogenides},
journal = {Advanced Science},
volume = {4},
number = {12},
pages = {1700231},
doi = {https://doi.org/10.1002/advs.201700231},
url = {https://advanced.onlinelibrary.wiley.com/doi/abs/10.1002/advs.201700231},
eprint = {https://advanced.onlinelibrary.wiley.com/doi/pdf/10.1002/advs.201700231},
year = {2017}
}

@article{chen_2022,
author = {Chen, Jiayao and Liu, Xiaojiang and Tian, Yujia and Zhu, Wei and Yan, Chunze and Shi, Yusheng and Kong, Ling Bing and Qi, Hang Jerry and Zhou, Kun},
title = {3D-Printed Anisotropic Polymer Materials for Functional Applications},
journal = {Advanced Materials},
volume = {34},
number = {5},
pages = {2102877},
doi = {https://doi.org/10.1002/adma.202102877},
url = {https://advanced.onlinelibrary.wiley.com/doi/abs/10.1002/adma.202102877},
eprint = {https://advanced.onlinelibrary.wiley.com/doi/pdf/10.1002/adma.202102877},
year = {2022}
}

@article{li_2021,
author = {Li, Xu and Liu, Haiyang and Ke, Congming and Tang, Weiqing and Liu, Mengyu and Huang, Feihong and Wu, Yaping and Wu, Zhiming and Kang, Junyong},
title = {Review of Anisotropic 2D Materials: Controlled Growth, Optical Anisotropy Modulation, and Photonic Applications},
journal = {Laser \& Photonics Reviews},
volume = {15},
number = {12},
pages = {2100322},
doi = {https://doi.org/10.1002/lpor.202100322},
url = {https://onlinelibrary.wiley.com/doi/abs/10.1002/lpor.202100322},
eprint = {https://onlinelibrary.wiley.com/doi/pdf/10.1002/lpor.202100322},
year = {2021}
}

@article{vinson-2011,
  title = {Bethe-Salpeter equation calculations of core excitation spectra},
  author = {Vinson, J. and Rehr, J. J. and Kas, J. J. and Shirley, E. L.},
  journal = {Phys. Rev. B},
  volume = {83},
  issue = {11},
  pages = {115106},
  numpages = {7},
  year = {2011},
  month = {Mar},
  publisher = {American Physical Society},
  doi = {10.1103/PhysRevB.83.115106},
  url = {https://link.aps.org/doi/10.1103/PhysRevB.83.115106}
}

@article{apra-2020,
	author = {Aprà, E. and Bylaska, E. J. and De Jong, W. A. and Govind, N. and Kowalski, K. and Straatsma, T. P. and Valiev, M. and Van Dam, H. J. J. and Alexeev, Y. and Anchell, J. and Anisimov, V. and Aquino, F. W. and Atta-Fynn, R. and Autschbach, J. and Bauman, N. P. and Becca, J. C. and Bernholdt, D. E. and Bhaskaran-Nair, K. and Bogatko, S. and Borowski, P. and Boschen, J. and Brabec, J. and Bruner, A. and Cauët, E. and Chen, Y. and Chuev, G. N. and Cramer, C. J. and Daily, J. and Deegan, M. J. O. and Dunning, T. H. and Dupuis, M. and Dyall, K. G. and Fann, G. I. and Fischer, S. A. and Fonari, A. and Früchtl, H. and Gagliardi, L. and Garza, J. and Gawande, N. and Ghosh, S. and Glaesemann, K. and Götz, A. W. and Hammond, J. and Helms, V. and Hermes, E. D. and Hirao, K. and Hirata, S. and Jacquelin, M. and Jensen, L. and Johnson, B. G. and Jónsson, H. and Kendall, R. A. and Klemm, M. and Kobayashi, R. and Konkov, V. and Krishnamoorthy, S. and Krishnan, M. and Lin, Z. and Lins, R. D. and Littlefield, R. J. and Logsdail, A. J. and Lopata, K. and Ma, W. and Marenich, A. V. and Del Campo, J. Martin and Mejia-Rodriguez, D. and Moore, J. E. and Mullin, J. M. and Nakajima, T. and Nascimento, D. R. and Nichols, J. A. and Nichols, P. J. and Nieplocha, J. and Otero-De-La-Roza, A. and Palmer, B. and Panyala, A. and Pirojsirikul, T. and Peng, B. and Peverati, R. and Pittner, J. and Pollack, L. and Richard, R. M. and Sadayappan, P. and Schatz, G. C. and Shelton, W. A. and Silverstein, D. W. and Smith, D. M. A. and Soares, T. A. and Song, D. and Swart, M. and Taylor, H. L. and Thomas, G. S. and Tipparaju, V. and Truhlar, D. G. and Tsemekhman, K. and Van Voorhis, T. and Vázquez-Mayagoitia, Á. and Verma, P. and Villa, O. and Vishnu, A.},
	journal = {The Journal of Chemical Physics},
	month = {5},
	number = {18},
	title = {{NWChem: Past, present, and future}},
	volume = {152},
	year = {2020},
	doi = {10.1063/5.0004997},
	url = {https://doi.org/10.1063/5.0004997},
}

@article{pople-1999,
	author = {Pople, John A.},
	journal = {Reviews of Modern Physics},
	month = {10},
	number = {5},
	pages = {1267--1274},
	title = {{Nobel Lecture: Quantum chemical models}},
	volume = {71},
	year = {1999},
	doi = {10.1103/revmodphys.71.1267},
	url = {https://doi.org/10.1103/revmodphys.71.1267},
}

@article{strobl-2007,
	author = {Strobl, Carolin and Boulesteix, Anne-Laure and Zeileis, Achim and Hothorn, Torsten},
	journal = {BMC Bioinformatics},
	month = {1},
	number = {1},
	title = {{Bias in random forest variable importance measures: Illustrations, sources and a solution}},
	volume = {8},
	year = {2007},
    pages = {25},
	doi = {10.1186/1471-2105-8-25},
	url = {https://doi.org/10.1186/1471-2105-8-25},
}

@article{abramson_x-ray_2025,
  author = {Abramson, Jared E. and Cardot, Charles A. and Kas, Josh J. and Rehr, John J. and Kaminsky, Werner and Michor, Herwig and Roman, Marta and Becker, Petra and Seidler, Gerald T.},
  journal = {The Journal of Physical Chemistry C},
  month = {11},
  title = {X-ray Emission Spectropolarimetry of Strongly Anisotropic Single Crystal Systems Using a Rowland Circle Geometry},
  year = {2025},
  volume = {129},
  number = {46},
  pages = {20649--20661},
  doi = {10.1021/acs.jpcc.5c05392},
  publisher = {American Chemical Society}
}

@book{molnar2025,
  title={Interpretable Machine Learning},
  subtitle={A Guide for Making Black Box Models Explainable},
  author={Christoph Molnar},
  year={2025},
  edition={3},
  isbn={978-3-911578-03-5},
  url={https://christophm.github.io/interpretable-ml-book}
}

@article{tortora-2024,
	author = {Tortora, L. and Tomassucci, G. and Pugliese, G. M. and Hacisalihoglu, M. Y. and Simonelli, L. and Marini, C. and Das, G. and Ishida, S. and Iyo, A. and Eisaki, H. and Mizokawa, T. and Saini, N. L.},
	journal = {Physical Chemistry Chemical Physics},
	month = {1},
	number = {34},
	pages = {22454--22462},
	title = {{Anisotropic atomic displacements, local orthorhombicity and anomalous local magnetic moment in Ba$_{0.6}$K$_{0.4}$Fe$_2$As$_2$ superconductor}},
	volume = {26},
	year = {2024},
	doi = {10.1039/d4cp02345e},
	url = {https://doi.org/10.1039/d4cp02345e},
}

@article{torrisi-2020,
	author = {Torrisi, Steven B. and Carbone, Matthew R. and Rohr, Brian A. and Montoya, Joseph H. and Ha, Yang and Yano, Junko and Suram, Santosh K. and Hung, Linda},
	journal = {NPJ Computational Materials},
	month = {7},
	number = {1},
	title = {{Random forest machine learning models for interpretable X-ray absorption near-edge structure spectrum-property relationships}},
	volume = {6},
	year = {2020},
    pages = {109},
	doi = {10.1038/s41524-020-00376-6},
	url = {https://doi.org/10.1038/s41524-020-00376-6},
}

@article{iwayama-2022,
	author = {Iwayama, Megumi and Wu, Stephen and Liu, Chang and Yoshida, Ryo},
	journal = {Journal of Chemical Information and Modeling},
	month = {10},
	number = {20},
	pages = {4837--4851},
	title = {{Functional Output regression for machine learning in materials science}},
	volume = {62},
	year = {2022},
	doi = {10.1021/acs.jcim.2c00626},
	url = {https://doi.org/10.1021/acs.jcim.2c00626},
}

@article{guda-2021,
	author = {Guda, A. A. and Guda, S. A. and Martini, A. and Kravtsova, A. N. and Algasov, A. and Bugaev, A. and Kubrin, S. P. and Guda, L. V. and Šot, P. and Van Bokhoven, J. A. and Copéret, C. and Soldatov, A. V.},
	journal = {NPJ Computational Materials},
	month = {12},
	number = {1},
	title = {{Understanding X-ray absorption spectra by means of descriptors and machine learning algorithms}},
	volume = {7},
	year = {2021},
    pages = {203},
	doi = {10.1038/s41524-021-00664-9},
	url = {https://doi.org/10.1038/s41524-021-00664-9},
}

@article{loudon-1964,
	author = {Loudon, R.},
	journal = {Advances In Physics},
	month = {10},
	number = {52},
	pages = {423--482},
	title = {{The Raman effect in crystals}},
	volume = {13},
	year = {1964},
	doi = {10.1080/00018736400101051},
	url = {https://doi.org/10.1080/00018736400101051},
}

@book{ziman-1972,
	author = {Ziman, J. M.},
	month = {7},
	title = {{Principles of the theory of solids}},
	year = {1972},
    publisher = {Cambridge University Press},
	doi = {10.1017/cbo9781139644075},
	url = {https://doi.org/10.1017/cbo9781139644075},
}

@article{cahill-2003,
	author = {Cahill, David G. and Ford, Wayne K. and Goodson, Kenneth E. and Mahan, Gerald D. and Majumdar, Arun and Maris, Humphrey J. and Merlin, Roberto and Phillpot, Simon R.},
	journal = {Journal of Applied Physics},
	month = {1},
	number = {2},
	pages = {793--818},
	title = {{Nanoscale thermal transport}},
	volume = {93},
	year = {2003},
	doi = {10.1063/1.1524305},
	url = {https://doi.org/10.1063/1.1524305},
}

@book{glassner-2021,
	author = {Glassner, Andrew},
	month = {6},
	publisher = {No Starch Press},
	title = {{Deep learning}},
	year = {2021},
}

\end{document}